\documentclass[11pt]{book}

\usepackage{amsmath,graphicx,amssymb}

\begin{document}

\setlength{\textheight}{8in} 
\setlength{\textwidth}{5.5in}
\setlength{\oddsidemargin}{0.5in}
\setlength{\evensidemargin}{0.5in}

\pagestyle{myheadings} 
\markboth{V.I. Marconi, D. Dom{\'i}nguez}{Non equilibrium phase diagrams of 
current driven JJA}
\renewcommand{\chaptername}{\vskip-20mm\protect\normalsize\protect\huge\bf Chapter}

\setcounter{chapter}{0}
\chapter{Non equilibrium phase diagrams of current driven Josephson junction arrays}
\setcounter{page}{1}

\begin{center}
{\Large {\bf  Ver\'onica I. Marconi$^\dag$, Daniel Dom{\'i}nguez$^\ddag$} } \\
\vspace{1.5cm}
{ \large {\it  $^\dag$ Institut de physique, Universit\'e de Neuch\^{a}tel,\\
 CH-2000 Neuch\^{a}tel,
 Switzerland.\\
  e-mail: veronica.marconi@unine.ch}}\\

\vspace{1.1cm}

{\large {\it  $^\ddag$Centro At\'{o}mico Bariloche,\\
 8400 S. C. de Bariloche, R{\'i}o Negro, Argentina.\\
 e-mail: domingd@cab.cnea.gov.ar}}\\

\begin{minipage}{5in}
\centerline{{\sc Abstract}}
\medskip

We  present a review  of our previous numerical studies on
non equilibrium  vortex dynamics in Josephson Junction arrays (JJA)
driven by a   dc current. Dynamical phase diagrams for different
magnetic fields, current directions
 and  varying temperature are discussed and compared.

First, the effect of thermal fluctuations in a current
driven diluted vortex lattice (VL)  is analyzed.
The case of  $f=1/25$ (where $f$ is the fraction of flux quanta
per plaquette in the array) is considered and
the phase diagram
as a function of the driving current and temperature is analyzed.
In equilibrium, this system has a weakly first-order melting transition
of the vortex lattice, which coincides with a depinning transition.
When a low current is applied,
the  ``longitudinal'' depinning transition occurs at a
temperature lower than the melting transition.  More interestingly, for
large currents (well above the critical current)
there is an analogous sequence of transitions but for
the transverse response of a fast moving VL. There
is a {\it transverse depinning temperature} below the melting
transition of the moving VL.

We also discuss the dependence with the direction of the applied dc 
current of
the transport properties of diluted vortex arrays on a  square JJA
at low temperatures.
  We show that  {\it
orientational pinning} phenomenon leads to a  finite transverse
critical current when the bias current  is applied in the  directions
of high symmetry and it leads to an {\it anomalous transverse voltage}
when vortices are  driven away from the favorable directions. In
addition, the effect of    disorder  in the transport
properties  of  square  JJA with a dc current applied in the ``diagonal
direction'' ([11] direction) is analyzed and a finite transverse voltage
is  also observed in this case.

The case of a fully frustrated square JJA, corresponding to $f=1/2$,
driven by a dc current and with  thermal fluctuations is also discussed.
In equilibrium, the low temperature phase has two broken symmetries:
the U(1) symmetry, corresponding to superconducting coherence, and
the $Z_2$ symmetry corresponding to the periodic order of the VL,
which forms a ``checkerboard pattern".
At high currents (well above the critical current)
two well separated transitions are observed.
The order of the checkerboard vortex lattice (discrete $Z_2$
symmetry) is destroyed at a much lower temperature than the transverse
superconducting coherence (continuous $U(1)$ symmetry).

\end{minipage}
\end{center}

\noindent {\bf Keywords:} Josephson, vortex,  phase transitions, non-equilibrium. 

\section{INTRODUCTION}

The behavior of superconducting vortices in the presence of periodic pinning shows very rich static and
dynamic phenomena \cite{martinoli,wire,jjarev,jjpin,jjafrus,dot,hole,sfield,yan,
martinoli0,PT,martinoli2,nelson,franz,hattel,nori0,carneiro0,nori,nori2,
carneiro99,reich,carneiro,prl1,prbnos,prb2,prl2}. The competition between the repulsive vortex-vortex interaction
and the attractive periodic pinning potential results in novel vortex structures at low temperatures
\cite{jjarev,sfield,yan,martinoli0,nori0,carneiro0}. The equilibrium phase transitions of these vortex
structures and their various dynamical regimes when driven out of equilibrium are of great interest
both experimentally \cite{martinoli,wire,jjarev,jjpin,jjafrus,dot,hole,sfield,yan,martinoli2}  and
theoretically \cite{jjarev,martinoli0,PT,martinoli2,nelson,franz,hattel,
nori0,carneiro0,nori,nori2,carneiro99,reich,carneiro,prl1,prbnos,prb2,prl2,jlt}.  Several techniques have been
developed to fabricate in superconducting samples an artificial periodic  pinning  structure: 
thickness modulated superconducting films \cite{martinoli}, superconducting wire networks \cite{wire}, 
Josephson junction arrays \cite{jjarev,jjpin,jjafrus}, magnetic dot arrays \cite{dot},  submicron hole
lattices \cite{hole,sfield}, and pinning induced by Bitter decoration \cite{yan}. The ground states of
these systems can be either commensurate or incommensurate vortex structures depending on the vortex
density (i.e the magnetic field). In the commensurate case, a ``matching'' field is defined  when the
number of vortices $N_v$ is an integer multiple of the number of pinning sites $N_p$: $N_v=nN_p$.  A
``submatching'' or ``fractional'' field is defined when  $N_v$ is a rational multiple of $N_p$:
$N_v=fN_p$ with $f=p/q$. One of the main properties of periodic pinning is that there are enhanced
critical currents and resistance minima both for fractional and for matching magnetic fields, for which
the vortex lattice is strongly pinned. 

In the case of JJA \cite{jjarev} the discrete lattice structure  of
Josephson junctions induces an effective periodic pinning potential (the so-called ``egg-carton''
potential) which at low temperatures confines the vortices at the centers of the unit cells of the
network \cite{jjpin}. There are strong commensurability effects for submatching fields $f=p/q$, for
which the vortices arrange in an ordered $q\times q$ superlattice that is commensurate with the
underlying array of junctions. The transition temperature $T_c(f)$ and the critical current $I_c(f)$
have maxima for  rational $f=p/q$, which have been observed experimentally \cite{jjarev,jjafrus}.
Moreover, as we will discuss here,  the model that describes the physics of the JJA can be thought as a
discrete lattice London model for thin film superconductors with periodic arrays of holes.  However,
this comparison can be valid only for low submatching fields since it can not describe the effects of
interstitial vortices.

The equilibrium  phase transitions at finite temperatures of two dimensional  systems with periodic
pinning  have been studied in the past \cite{PT,martinoli2,nelson,franz,hattel}. It is possible to
have a depinning phase transition of the commensurate ground states at a temperature $ T_{p} $ and a
melting transition of the vortex lattice at a temperature $ T_{M} $ 
\cite{PT,martinoli2,nelson,franz,hattel}.  Franz and  Teitel \cite{franz}  have studied this problem for
the case of submatching fields. For $0<T<T_p$ there is  a pinned phase in which the vortex lattice (VL)
is pinned commensurably to the periodic potential and has long-range order. For $T_p<T<T_M$ there is a
floating VL which is depinned and has quasi-long-range order.  For high submatching fields ($f\gtrsim
1/30$ ) both transitions coincide, \( T_{p}=T_{M} \), while for low submatching fields ($f\lesssim
1/30$)  both transitions are different with \( T_{p}<T_{M} \) \cite{franz,hattel}. 

The non-equilibrium dynamics of driven  vortex lattices interacting either with random or periodic
pinning shows  an interesting  variety of behavior \cite{nori,nori2,carneiro99,reich,carneiro,prl1,prbnos,prb2,prl2,
KV,gld,bmr,pardo,simu,dgb,dd99}.  Many recent studies have concentrated in the  problem of the
driven VL in the presence of {\it random} pinning  \cite{KV,gld,bmr,pardo,simu,dgb,dd99}. When there is
a large driving current the effect of the pinning potential is reduced, and the nature of the fastly
moving vortex structure has been under active discussion. The moving vortex phase has been proposed to
be either a crystalline structure, a moving glass,  a moving smectic or a moving transverse glass
\cite{KV,gld,bmr}. These moving phases have been studied both experimentally \cite{pardo} and
numerically \cite{simu,dgb,dd99}.  Motivated by these results, the dynamical regimes of  the moving VL
in the presence of {\it periodic} pinning has also become a subject of
interest \cite{nori,nori2,carneiro99,reich,carneiro,prl1, prbnos, prb2,prl2,jlt}. At zero temperature,  the dynamical
phases of vortices driven by an external current with a periodic array of pinning sites has been
studied in very detail by F. Nori and coworkers \cite{nori,nori2}. A complex variety of regimes has been
found, particularly for $N_v>N_p$ where the motion of interstitial vortices leads to several
interesting dynamical phases.

Moreover, most of the effects of periodic pinning that have been studied are related
to conmensurability phenomena and the breaking of translational symmetry in these
systems. Less studied is the effect of the breaking of rotational symmetry
in periodic pinning potentials, in particular regarding transport
properties. One question of interest is how the motion of vortices changes
when the direction of the driving current is varied. If there is rotational
symmetry, the vortex motion and voltage response should be insensitive to
the choice of the direction of the current. However, it is clear that in a
periodic pinning potential the dynamics may depend on the direction of the
current. For example, in square JJA it has been
found experimentally 
that the existence of fractional giant Shapiro steps (FGGS) depends on
the orientation of the current bias. When the JJA is driven in the [11]
direction the FGGS are absent, while they are very large when the drive
is in the [10] direction \cite{sohn}. 
But may be the another  example of more recent interest
is the phenomenon of transverse critical current in superconductors with
pinning, already mentioned before \cite{KV,gld,bmr}. Eventhough many numerical results
 have  shown the existence of a critical transverse current \cite{nori, nori2, prl1}
 there has not been experimental measurements 
of the existence of transverse voltages and transverse pinning 
effects before our initial experimental-numerical work \cite{prbnos}.  

In particular, a very interesting case of conmesurability effect in JJA, where 
the non-equilibrium vortex dynamics could be study are the fully frustrated JJA.
In the presence of a magnetic field such that there is
a half flux quantum per plaquette, $f=Ha^2/\Phi_0=1/2$, the JJA corresponds
to the fully frustrated XY (FFXY) model \cite{teitel83,berge86,expff,teitel89}. 
The ground state is
 a ``checkerboard'' vortex lattice, in which a vortex sits
in every other site of an square grid \cite{teitel83}.
 There are two types of competing
order and broken symmetries: the discrete
$Z_2$ symmetry of the ground state of the vortex lattice,
with an associated chiral (Ising-like) order parameter, 
and the continuous $U(1)$ symmetry associated with superconducting
phase coherence. 
The critical behavior of this system has been the
subject of several experimental \cite{expff} and theoretical 
\cite{teitel83,berge86,teitel89,granato91,ramirez92,olson95,granato97,diep98}
studies.
There are a  $Z_2$ transition (Ising-like) and 
a $U(1)$ transition (Kosterlitz-Thouless-like) with critical
temperatures  $T_{Z_2}\ge T_{U(1)}$. There is a controversy
about these temperatures being extremely close \cite{olson95}
or equal \cite{ramirez92,granato97,diep98}. 
  In  the light of this, it is worth studying the possibility
of non-equilibrium $Z_2$ and $U(1)$ transitions at large driving currents. 
Also, the dynamical transitions in driven systems studied up to now
involve continous (translational or gauge) symmetries, 
and therefore it is interesting to study a system with
a discrete symmetry.

In this review we collect our previous works \cite{prl1, prbnos, prb2, prl2, jlt}  on 
the  study of the dynamical regimes  of a moving VL in  the periodic pinning of a  Josephson
junction array (JJA) of $L$ x $L$ junctions,  at {\it finite temperatures} for different cases of  {\it submatching}
 field
($f=1/25$, $f=1/L^2$ and  $f=1/2$).  For the case of diluted vortex lattices ($f=1/25$)
 we obtain a  phase diagram as a function of the driving current $I$ and temperature $T$.
 We find that when the VL is driven by a low current, the depinning and
melting transitions can become separated even for a field for which they coincide in equilibrium.
Moreover, we can distinguish between the depinning of the VL in the direction of the current drive, and
the \textit{transverse depinning} in the direction perpendicular to the drive. This later case
corresponds to the vanishing of the transverse critical current in a moving VL at a given temperature
$T_{tr}$, or equivalently, to the vanishing of the transverse superconducting coherence.  We obtain
three distinct regimes at low temperatures: (i){\it Pinned vortex lattice:} for $0<T<T_p(I)$ there is
an ordered VL which has crystaline long-range order, superconducting coherence (i.e., a finite helicity
modulus)  and zero resistance  both in the longitudinal and transverse directions. (ii){\it
Transversely pinned vortex lattice:} for $T_p(I)<T<T_{tr}(I)$ there is a moving VL which has
anisotropic Bragg peaks, quasi-long range order, transverse superconducting coherence and  zero
transverse resistivity. There is a finite transverse critical current. This regime also has strong
orientational pinning effects\cite{prbnos}  in the [1,0] and [0,1] lattice directions. (iii){\it
Floating vortex lattice:} for $T_{tr}(I)<T<T_M(I)$ there is a moving VL which is unpinned in both
directions and it has quasi-long range crystalline order with a strong anisotropy.  After
our first work in Ref.~\cite{prl1},  further studies of thermal effects in a moving vortex lattice with a
periodic pinning array have  been reported for $N_v=N_p$ \cite{reich}  and  for $N_v>N_p$
\cite{carneiro99,carneiro}. Some of these results are similar to ours.

Regarding the breaking of rotational invariance
in square JJA, case in which  
the discrete lattice of Josephson junctions induces a
periodic egg-carton potential for the motion of vortices \cite{jjpin}, 
we will show 
how the voltage response depends on the angle of the current with respect to
the lattice directions of the square JJA. We will show  numerically that there are
preferred directions for vortex motion for which there is orientational
pinning. These results are in good agreement with experimental results.
 This leads to an anomalous transverse voltage when vortices are
driven in directions different from the symmetry directions. An analogous
effect of a transverse voltage due to the guided motion of vortices has been
observed experimentally 
in YBCO superconductors with twin boundaries \cite{twins}.

As we mentioned before, in a first
 stage of our research we have found  dynamical transitions of the vortex lattice in a JJA 
with a field density of $f=1/25$ 
\cite{prl1,prb2}:  for large currents $I$ 
there is a melting transition of the moving vortex lattice
at a temperature higher than the transverse superconducting
transition: $T_M(I)>T_{U(1)}(I)$. 
Interestingly, in the fully frustrated case ($f=1/2$) we find that the opposite
case occurs in the driven FFXY: 
the order of the ``checkerboard'' vortex
lattice is destroyed at a much lower temperature than the
transverse superconducting coherence, $T_{Z_2}(I)<T_{U(1)}(I)$.

The remainder of our review is organized as follows.
In Sec.~2 we introduce the theoretical model used for the dynamics
of the JJA. We also discuss how  this model can be mapped to a superconducting
film with a periodic array of holes. 
In Sec.~3 we discuss the results for the case of diluted vortex lattices dynamics in JJA
including the orientational pinning results. 
In Sec.~4 we present the results on the
  non-equilibrium dynamical regimes in fully frustrated JJA.
In Sec.~5 we present   a discussion comparing the two previous sections and the general 
conclusions.   We also provide  a detailed definition
of the adequate periodic boundary conditions for a JJA with an external
magnetic field and an external driving current in Appendix A, as well as the
algorithm used for the numerical simulation in Appendix B.

\section{MODEL}

\subsection{Resistively Shunted Junction dynamics}


\begin{figure}[t]
\begin{center}
\begin{minipage}[h]{80mm}
\centerline{\includegraphics[width=8.0cm]{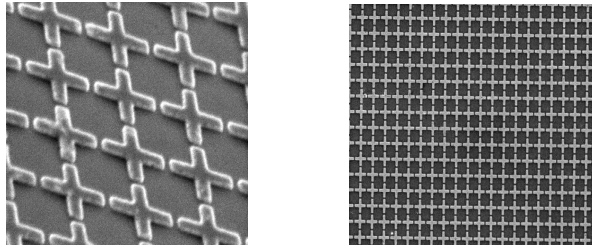}}
\end{minipage}
\caption{Experimental samples: square proximity-effect Pb/Cu/Pb Josephson Junction Arrays SEM's image, with an array
lattice space $a=10\mu m$ and distance between superconducting islands $d=1\mu m$ (the authors thank to Dr. H. Pastoriza and collaborators, Centro At\'omico Bariloche, Argentina, 
    for this picture.)} \label{fig0}
\end{center}
\end{figure}

We study the dynamics of square Josephson junction array (JJA) with $L\times L$ superconducting
nodes (see Fig.\ref{fig0}) using the Resistively Shunted Junction (RSJ) model for the
junctions \cite{dyna0,falo,eik,acvs,kim93}. The   nodes are in the lattice  sites ${\bf
n}=(n_{x},n_{y})$ and their  superconducting phases are $\theta ({\bf n})$. 
Basically we have an XY model, therefore its Hamiltonian  is the following:
\begin{equation}
{\cal H} = -\sum_{\mu,{\bf n}}\frac{\Phi_0 I_0}{2\pi}
\cos[\theta({\bf n}+{\bf \mu})-\theta({\bf n})-A_{\mu}({\bf n})]\;.
\end{equation}

Using  this model, 
the current flowing in the junction between two superconducting nodes in the JJA is modeled as
the sum of the Josephson supercurrent and the normal current:
\begin{equation} 
I_{\mu }({\bf n})=I_{0}\sin \theta _{\mu }({\bf n})+\frac{\hbar}{2e R_{N}}\frac{d\theta_{\mu}({\bf n})}{dt}+\eta _{\mu }({\bf  n},t) 
\label{rsj} 
\end{equation}
 being $I_{0}$  the critical current of the junction between the sites 
${\bf n}$ and ${\bf n}+{\bf \mu }$, 
 (${\bf \mu }={\bf \hat{x}},{\bf \hat{y}}$), $R_{N}$  the normal state resistance and
\begin{equation} 
 \theta_{\mu }({\bf n})=\theta ({\bf n}+{\bf \mu })-\theta ({\bf n})-A_{\mu }({\bf n})=\Delta _{\mu}\theta ({\bf n})-A_{\mu }({\bf n})
\end{equation}
the gauge invariant phase difference with 
\begin{equation}
A_{\mu }({\bf n})=\frac{2\pi }{\Phi _{0}}\int_{{\bf n}a}^{({\bf n}+{\bf \mu })a}
{\bf A}\cdot d{\bf l}.
\end{equation} 
The thermal noise fluctuations $\eta _{\mu }$ have correlations 
\begin{equation}
  \langle \eta _{\mu }({\bf n},t)\eta _{\mu ^{\prime }}({\bf
n^{\prime }} ,t^{\prime })\rangle =\frac{2kT}{R_{N}}\delta _{\mu ,\mu ^{\prime }}\delta _{
{\bf n},{\bf n^{\prime }}}\delta (t-t^{\prime }).
\end{equation}

In the presence of an external magnetic field $H$ we have
\begin{eqnarray}
\Delta_{\mu}\times A_{\mu}({\bf n})
   &=&A_x({\bf n})-A_x({\bf n}+{\bf y})+  A_y({\bf n}+{\bf x})-A_y({\bf n})\nonumber\\ 
   &=&2\pi f, 
\end{eqnarray}
    $f=H a^2/\Phi_0$ and $a$ is the array lattice spacing. If the magnetic field is such that  $f=1/L^2$ it
corresponds to have one vortex in the sample,  if $f=1/25$ we have an example of diluted vortex lattices, if $f=1/2$,
there is half flux quantum per plaquette and the JJA correspond to the fully frustrated XY model.

   We take periodic boundary conditions (PBC) in both directions in the presence of an external current ${\bf I}=I{\bf
\hat{y}}$ \cite{dd99} (See Appendix A). The vector potential is taken as
\begin{equation} 
  A_{\mu}({\bf n},t)=A_{\mu}^0({\bf n})-\alpha_{\mu}(t)
\end{equation}
 where in the Landau gauge $A^0_x({\bf n})=-2\pi f n_y$, $A^0_y({\bf n})=0$ and
$\alpha_{\mu}(t)$ allows for total voltage fluctuations under periodic boundary conditions.  In
this gauge the PBC for the phases are \cite{prl1,prb2,dd99}:
\begin{eqnarray}
\theta(n_x+L,n_y)&=&\theta(n_x,n_y)  \nonumber \\
\theta(n_x,n_y+L)&=&\theta(n_x,n_y)-2\pi f Ln_x.
\label{pbcp}
\end{eqnarray}

 We also consider local conservation of current,
\begin{equation}
\Delta_\mu\cdot I_{\mu}({\bf n})=\sum_{\mu} I_{\mu}({\bf n})- I_{\mu}({\bf n} -{\bf \mu})=0.
\label{divI} 
\end{equation}
 After Eqs.~(\ref{rsj},\ref{divI})  we obtain the following 
equations for the phases \cite{prl1,prb2,dd99},
\begin{equation}
\frac{\hbar}{2eR_{N}}\Delta_{\mu}^2\frac{d\theta({\bf n})}{dt}= -\Delta_{\mu}\cdot [S_{\mu}({\bf
n})+\eta_{\mu}({\bf n},t)]
\label{dyn0}
\end{equation} where
\begin{equation}
S_{\mu}({\bf n})=I_0\sin[\Delta_\mu\theta({\bf n})-A_{\mu}^0({\bf n})- \alpha_{\mu}]\;, \end{equation} and
the discrete Laplacian is
\begin{eqnarray}
\Delta^2_\mu f({\bf n})&=&f({\bf n}+{\bf \hat x})
+f({\bf n}- {\bf \hat x})+f({\bf n}+{\bf \hat y}) +f({\bf n}-{\bf \hat y} )\nonumber\\ &
&\;-4\,f({\bf n}). 
\end{eqnarray}
 The Laplacian can be inverted with the square lattice Green's
function $G_{{\bf n},{\bf n'}}$:
\begin{equation}
\Delta^2_\mu G_{{\bf n},{\bf n'}}=\delta_{{\bf n},{\bf n'}}.
\label{green}
\end{equation}

Since we take PBC (see Appendix A), the  total current has to be fixed by:

\begin{eqnarray}
 I_x&=&\frac{1}{L^2}\left[\sum_{ {\bf n}} I_0\sin\theta_x({\bf n})+\eta_x({\bf n},t)\right] + \frac{\hbar}{2eR_N} \frac{d\alpha_x}{dt}\;,  \nonumber \\ & &\label{tot}\\
I_y&=&\frac{1}{L^2}\left[\sum_{ {\bf n}} I_0 \sin\theta_y ({\bf n})+\eta_y({\bf n},t)\right] + \frac{\hbar}{ 2eR_N} \frac{d\alpha_y}{dt}\;,\nonumber
\end{eqnarray}

These
equations determine the dynamics of $\alpha_\mu(t)$ \cite{dd99}.  For the case of a current flowing in the $y$ direction   we take $I_x=0$
and $I_y=I$ and for the case  in which we apply a current $I$ at an angle $\phi$ with respect to the $[10]$
lattice direction, 
$I_x=I\cos\phi$ and   $I_y=I\sin\phi.$ After Eqs.~(\ref{dyn0},\ref{green},\ref{tot}) we obtain  the following set of
dynamical equations \cite{prl1,prb2,dd99},  

\begin{eqnarray}
\frac{d\theta({\bf n})}{dt}&=&-\sum_{\bf n'}G_{{\bf n},{\bf n'}}\Delta_{\mu} \cdot \left[S_{\mu}({\bf n'})+\eta_{\mu}({\bf n'},t)\right],
\label{dyn1} \\
\frac{d\alpha_{\mu}}{dt}&=& I_{ \mu } -\frac{1}{L^2}\sum_{{\bf n }}
S_{\mu}({\bf n})+\eta_{\mu}({\bf n},t),
\label{dyn2} 
\end{eqnarray}

where we have normalized
currents by $I_0$,  time by $\tau_J=2eR_{N}I_0/\hbar$,  and temperature by $I_0\Phi_0/2\pi
k_B$.

\subsection{Comparison with thin film with a periodic array of holes}

Let us consider a superconducting thin film with a square array of
holes, which act as pinning sites for vortices. There are $N_p=L^2$
pinning sites separated by a distance $a$. 
The current density in the superconducting film is given by
 the sum of the supercurrent and the normal current:

\begin{eqnarray}
{\bf J}&=&{\bf J}_S+{\bf J}_N\nonumber\\
{\bf J}&=&\frac{ie\hbar}{m^*}
\left[\Psi^*{\bf D}\Psi-({\bf D}\Psi)^*\Psi\right]+
\frac{\sigma\Phi_0}{2\pi c}\frac{\partial}{\partial
t}\left(\nabla\theta-\frac{2\pi}{\Phi_0}{\bf A}\right),\nonumber\\
&&
\end{eqnarray}
with ${\bf D}=\nabla+i\frac{2\pi}{\Phi_0}{\bf A}$, 
$\Psi({\bf r})=|\Psi({\bf r})|\exp[i\theta({\bf r})]$  the
superconducting order parameter and $\sigma$  the normal state
conductivity. These equations are valid everywhere
in the film except in the hole regions. If the number of vortices
$N_v=BL^2a^2/\Phi_0$ is much smaller than the number of pinning sites
$N_p$, all vortices will be centered in the holes in equilibrium. 
In this case we can assume that $|\Psi({\bf
n})|\approx|\Psi_0|$  is homogeneous in the superconducting film.
Therefore the dynamics is given by the superconducting phase 
$\theta({\bf r})$,  corresponding to a London model in a sample with holes.
After considering current conservation
$ \nabla\cdot{\bf J}=0,$ we obtain the London dynamical equations 
for the phases
in this multiply connected geometry. Since $N_v\ll N_p$, we make the 
approximation of solving the equations in a discrete grid of spacing $a$.
This means that we take as the relevant dynamical variables the phases 
$\theta({\bf r}_{\bf n})$ defined
in the sites which are dual to the pinning sites. They represent
the average superconducting phase in each
superconducting square defined by four pinning sites. 
Therefore, we  take the discretization ${\bf r}_{\bf
n}=(n_xa,n_ya)=a{\bf n}$. [Pinning sites at centered
at positions ${\bf r}_{\bf p}=(n_x + 1/2 , n_y+1/2)a$]. 
The derivatives in the supercurrent are discretized in a gauge-invariant way as
\begin{equation}
D_\mu \Psi({\bf r})\rightarrow\frac{1}{a}\left[\Psi({\bf n}+
{\bf\mu})-e^{-i2\pi A_\mu({\bf n})/\Phi_0}\Psi({\bf n})\right].
\end{equation}
After doing this, we obtain an equation analogous to (\ref{rsj}).
 Now $I_\mu({\bf n})$ has to be interpreted as current density
normalized by $J_0=2e\hbar |\Psi_0|^2/ma=\Phi_0/(8\pi^2\lambda^2a)$,
time normalized by $\tau=c/(4\pi\sigma\lambda^2)$,
and the fraction of vortices is  $f=N_v/N_p=Ba^2/\Phi_0$.
This leads to a set of dynamical equations of the same form as 
Eqs.(\ref{dyn1},\ref{dyn2}). Therefore, we expect that for $f\ll 1$ the
model for a JJA also gives a good representation of the physics
of a superconducting film with a square array of holes
(meaning that effects of interstitial vortices are neglected
for $N_v\ll N_p$). In other words, we expect that for a low
density of vortices the specific shape of the periodic pinning
potential (being either an egg-carton  or an  array of holes)
will not be physically relevant.

\subsection{Quantities calculated and simulation parameters}

The Langevin dynamical equations (\ref{dyn1},\ref{dyn2}) 
are solved with a second order
Runge-Kutta-Helfand-Greenside algorithm with time step $\Delta
t=(0.05-0.1)\tau_J$. The discrete periodic Laplacian is inverted with 
a fast Fourier + tridiagonalization algorithm 
(see the Appendix B for a detail of the algorithm used).

The main physical quantities calculated are the following:

(i) {\it Transverse superconducting coherence}: We obtain the helicity modulus
$\Upsilon_x$ in the direction transverse to the current as
\begin{eqnarray}
\Upsilon_x&=&\frac{1}{L^2}\left\langle\sum_{{\bf n}}\cos\theta_x({\bf
n})\right\rangle-\frac{1}{TL^4}\left\{\left\langle \left[\sum_{{\bf n}}
\sin\theta_x({\bf n})\right]^2\right\rangle\right. \nonumber\\ 
& & \left.- \left\langle \left[\sum_{{\bf n}}\sin\theta_x({\bf
 n})\right]\right\rangle^2\right\}\nonumber
\end{eqnarray}
Whenever we calculate the helicity modulus along $x$, we enforce
strict periodicity in $\theta$ by fixing $\alpha_x(t)=0$  
(see Appendix A).
 
(ii){\it Transport}: We calculate the transport response of the JJA
 from the time average of the total voltage as 
\begin{eqnarray}
V_x&=&\langle v_x(t)\rangle= \langle d\alpha_x(t)/dt\rangle  \nonumber \\
V_y&=&\langle v_y(t)\rangle= \langle d\alpha_y(t)/dt\rangle
\end{eqnarray}
with voltages normalized by $R_ {N}I_0$.

\begin{figure}[t]
\begin{center}
\begin{minipage}[h]{80mm}
\centerline{\includegraphics[width=8.0cm]{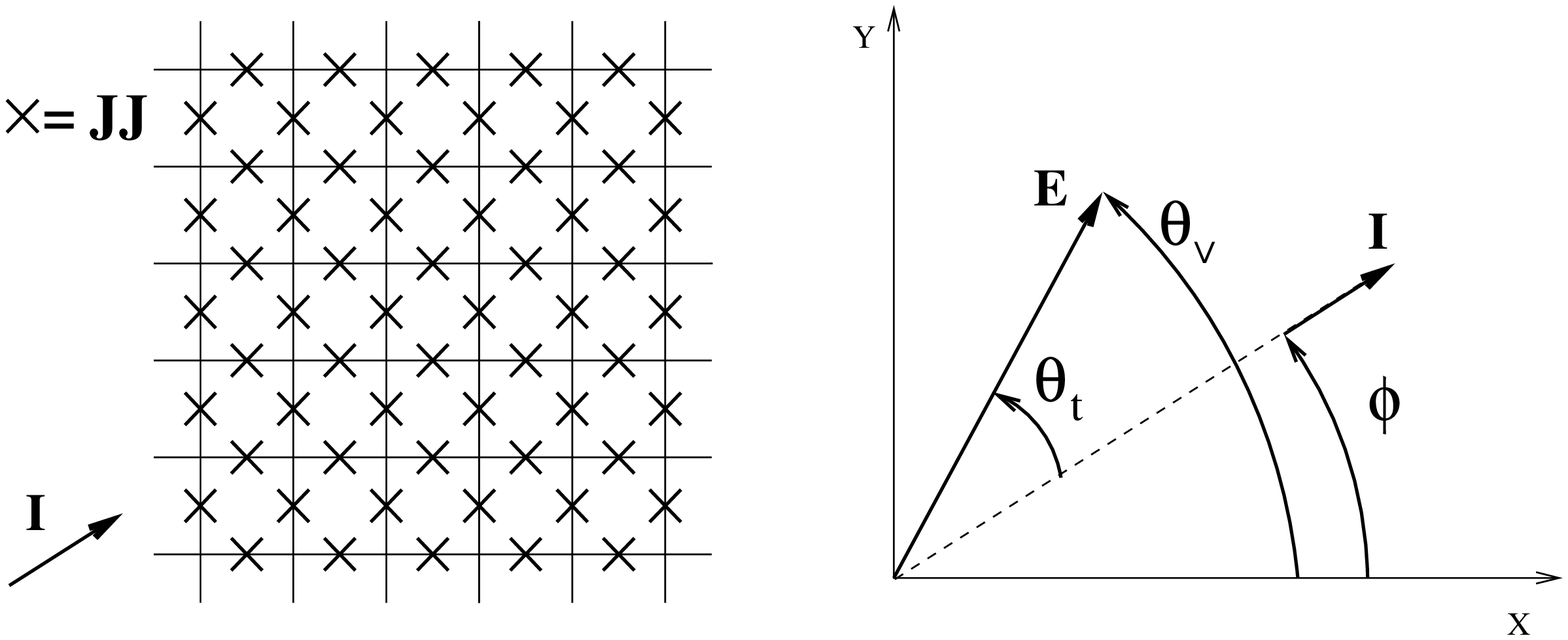}}
\end{minipage}
\caption{Left: schematic representation of a square Josephson junction
array, the crosses indicate Josephson junctions (JJ), with a rotated applied external current.
Right: definition of angles with respect to the $x$ and $y$ axis of
the array. The vector of applied current ${\bf I}=(I_x,I_y)$ forms an  angle
$\phi$ respect to the $x$-axis. The vector of electric field ${\bf E}=(V_x,V_y)$
forms an angle $\theta_v$ with respect to the $x$-axis and an angle
$\theta_t$ with respect to the current ${\bf I}$.} \label{fig1}
\end{center}
\end{figure}

(iii) {\it Vortex structure}: We obtain the vorticity
at the plaquette ${\bf \tilde n}=(n_x+1/2,n_y+1/2)$ 
(associated to the site ${\bf n}$) as \cite{dj96}: 
\begin{equation}
b({\bf \tilde n})=-\Delta_\mu\times{\rm nint}[\theta_\mu({\bf n})/2\pi]
\end{equation}
with ${\rm nint}[x]$ the nearest integer of $x$. We calculate the average
vortex structure factor as
\begin{equation}
S({\bf k})=\left\langle\left|\frac{1}{L^2}\sum_{\bf \tilde n}
b({\bf \tilde n})\exp(i{\bf k}\cdot{\bf \tilde n})\right|^2\right\rangle.
\end{equation}

Analyzing above properties we study  JJA under different magnetic fields applied. (a) Diluted vortex lattices corresponding to $f=1/25$. (b) 
A single vortex in
the array, $f=1/L^2$. (c) And  fully frustrated JJA, corresponding to have half 
flux quantum per plaquette, {\it i.e.} $f=1/2$.  We consider square networks of  $L\times L$ junctions, with 
$L=8, 16, 24, 32, 48, 64, 128$ or $L=50,100,150$. We  apply dc currents in different directions: 
(i) $I$ flowing in the $y$ direction or  (ii) $I$ at an angle $\phi$ with respect to the $[10]$
lattice direction, 
\begin{eqnarray}
I_x&=&I\cos\phi  \nonumber \\
I_y&=&I\sin\phi.
\end{eqnarray}
We define the longitudinal voltage as the voltage in the direction of the
applied current, 
\begin{equation}
V_l=V_x\cos\phi+V_y\sin\phi,
\label{lv}
\end{equation}
and the transverse voltage 
\begin{equation}
V_t=-V_x\sin\phi+V_y\cos\phi.
\label{tv}
\end{equation}
From the voltage response, we define the transverse angle as 
\begin{equation}
\tan\theta_t=V_t/V_l
\end{equation}
and the voltage angle as 
\begin{equation}
\tan\theta_v=V_y/V_x,
\end{equation}
i.e. $\theta_t=\theta_v-\phi$, see Fig.~\ref{fig1}. 
When the vortices move in the direction
perpendicular to the current, there is no transverse voltage, therefore $\theta_t=0$ and $\theta_v=\phi$.
The case with a dc current applied in the $y$ direction, ${\bf I} = I{\bf \hat y}$ ($\phi=\pi/2$), will be the more extensively used in this review.

%

\section{DILUTED VORTEX LATTICE DYNAMICS}

In this section we show the   study of JJA with a low magnetic field
  corresponding to $f=1/25$, diluted vortex lattices,  for
different system sizes of $L\times L$ junctions, with $L=50,100,150$.
Most of the results are for $L=100$, except when it is explicitly specified, 
and for $N_t=10^5$ iterations after a transient of $N_t/2$ iterations.

\subsection{Transition near equilibrium}

\begin{figure}[tbp]
\begin{center}
\begin{minipage}[h]{55mm}
\centerline{\includegraphics[width=6.0cm]{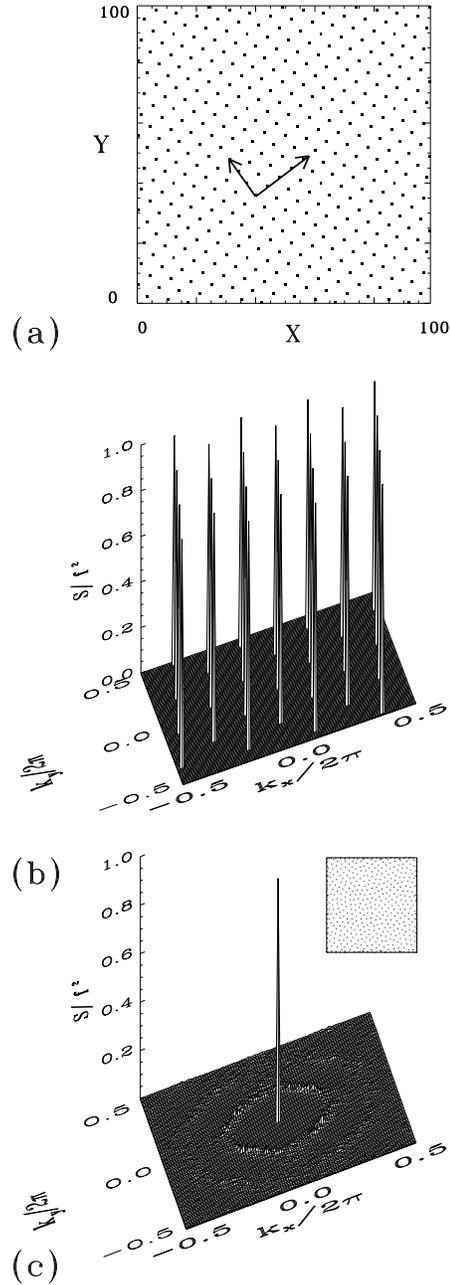}}
\end{minipage}
\caption{Vortex configuration for a low current $I=0.01 \ll I_c(0)$ 
and a low temperature, $T=0.01 < T_M$: 
(a) tilted square vortex lattice, oriented in the $[4a,3a]$ direction
and (b) corresponding structure factor $S({\bf k})$.
At high temperature, $T=0.05 \gtrsim T_M$:
(c) disordered vortex array. Inset: vortex liquid-like structure factor.} \label{fig2}
\end{center}
\end{figure}

\begin{figure}[t]
\begin{center}
\begin{minipage}[h]{80mm}
\centerline{\includegraphics[width=8.0cm]{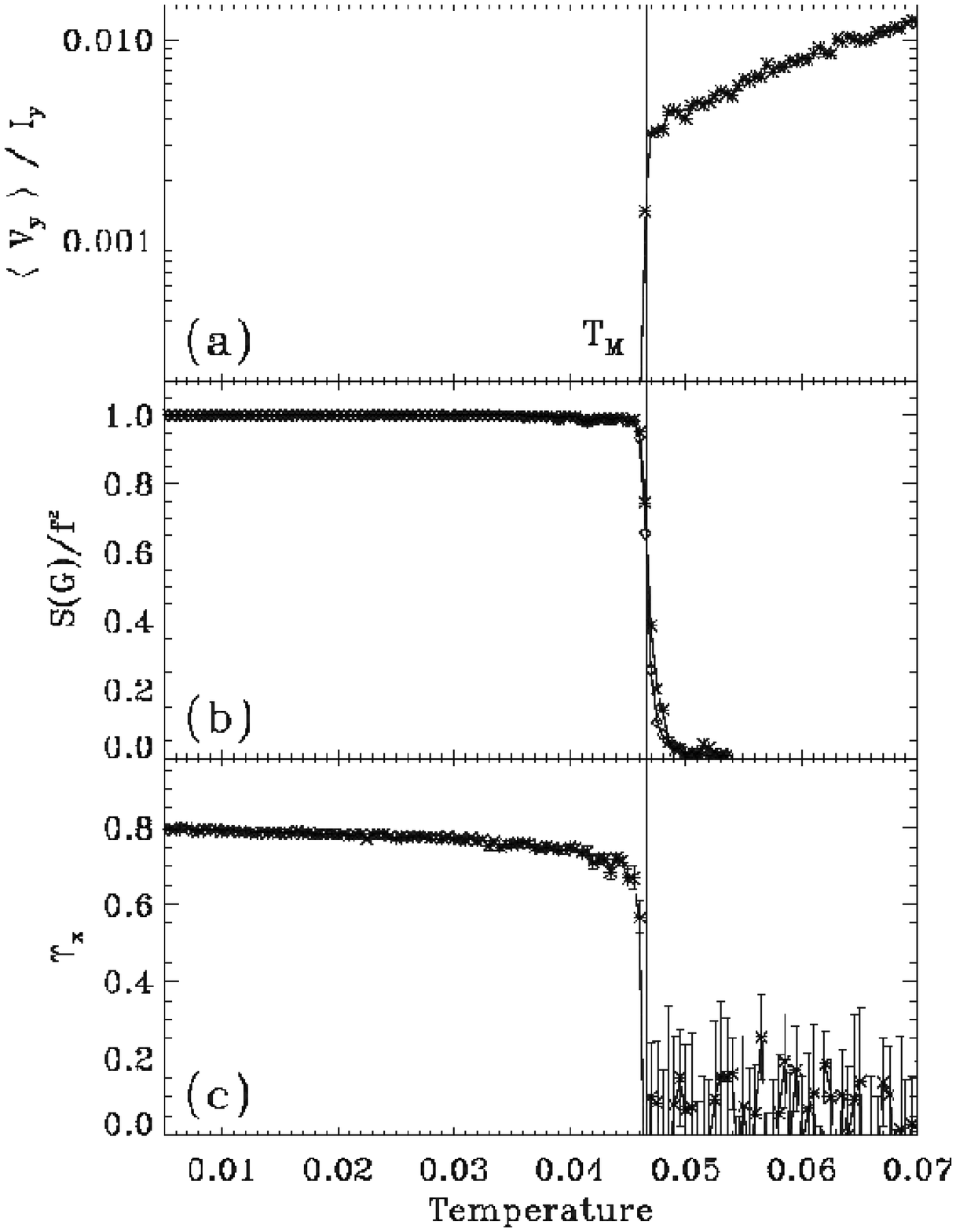}}
\end{minipage}
\caption{For low current $I \ll I_c(0)$, $I=0.01$:
(a) $<V_y>/I_y$ vs. $T$,
(b) $S({\bf G_1})$ $(\diamond)$ and $S({\bf G_2})$ $(\star)$ vs. $T$,
(c) $\Upsilon_x$ vs. $T$.
} \label{fig3}
\end{center}
\end{figure}

The ground state vortex configuration for $f=1/25$  is a tilted 
square-like vortex lattice (VL) \cite{teitel25}, see Fig.\ref{fig2}(a). We find
that this state
is stable for low currents and low temperatures (in fact, the structure
of Fig.\ref{fig2}(a) corresponds to $I=0.01$ and $T=0.01$). The lattice is
oriented in the $[4a,3a]$ direction and commensurated with the underlying
periodic pinning potential of the square JJA.
The structure factor $S({\bf k})$ has the corresponding Bragg
peaks at wavevectors ${\bf G}$ in the reciprocal space, 
as can be seen in Fig.\ref{fig2}(b). 
When the temperature is increased, the VL tends to disorder and above
the melting temperature
$T_M$ a random vortex array with
a liquid-like structure factor is obtained, Figs.\ref{fig2}(c) and inset.

We find a single equilibrium phase transition ($I=0$)
at $T_M\approx0.050\pm0.003$, which is in agreement with
the melting temperature obtained by Franz and Teitel\cite{franz} 
and Hattel and Wheatley\cite{hattel} for $f\gtrsim 1/30$.

We now apply a  very low current,  $I=0.01$, in order to
study the near-equilibrium transport response simultaneously with other
quantities like structure factor and helicity modulus.   We find a
phase transition at a temperature $T_M(I)\approx0.046\pm0.001$, which
is slightly lower than the equilibrium transition. In Fig.\ref{fig3}(a) we see
that there is a large jump in the resistance $R=V/I$  at $T_c$, in good
agreement with the first-order nature of the equilibrium transition
\cite{franz}. 
The onset of resistivity is a signature of a 
\begin{figure}[here]
\begin{center}
\begin{minipage}[h]{80mm}
\centerline{\includegraphics[width=8.0cm]{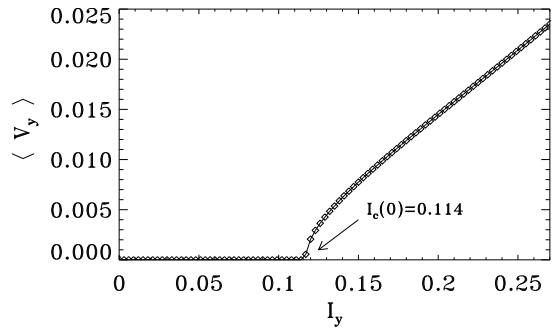}}
\end{minipage}
\caption{Zero temperature  $IV$ curve. There is a 
critical current $I_c(0)=0.114\pm0.002$.
} 
\label{fig4}
\end{center}
\end{figure}
\clearpage
depinning
transition in the direction of the drive.  This occurs simultaneously
with a melting of the vortex lattice, corresponding to the vanishing
of Bragg peaks,  as shown in Fig.\ref{fig3}(b) for the two first reciprocal lattice
vectors  ${\bf G_1}=\frac{2\pi}{a}(-4/25,-3/25)$ and ${\bf
G_2}=\frac{2\pi}{a}(-3/25,4/25)$.
  In the direction of the current drive the helicity modulus
is ill-defined since total phase fluctuations are allowed
(see Appendix A).  However, in
the perpendicular direction to the drive the helicity modulus
$\Upsilon_x$ can be calculated, and it is a measure of the transverse
superconducting coherence. As we can see in Fig.\ref{fig3}(c), transverse
superconductivity also vanishes at $T_M(I)$. Above $T_M$, we find that
the $\Upsilon_x(T)$ has large fluctuations around zero. 
\begin{figure}[t]
\begin{center}
\begin{minipage}[h]{80mm}
\centerline{\includegraphics[width=8.0cm]{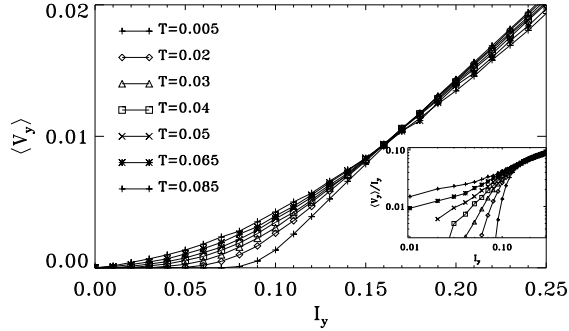}}
\end{minipage}
\caption
{$IV$ curves for different temperatures.
 Linear scale, crossing point at $I^*=0.165$.
Inset: Log-log plot of $R=V/I$ vs. $I$ for the same temperatures.  
}
\label{fig5}
\end{center}
\end{figure}

\subsection{Transport properties}

Let us now study the transport properties  
for larger currents. We calculate the current-voltage (IV)
characteristics for  different temperatures as well as the dc resistance
$R=V/I$ as a function of temperature for finite currents.
 
The zero temperature IV curve has a critical current, $I_c(0)=
0.114\pm0.002$, see Fig.\ref{fig4}, which corresponds to the single vortex
depinning current in square JJA \cite{jjpin}, with the typical square
root depence at the onset. Similar
behavior has been reported for zero
temperature IV curves for low values of $f$ \cite{vorstroud,vorjose}.  
Above $I_c(0)$ there is an
almost linear  increase of voltage, corresponding to a ``flux flow''
regime, where there is a fastly moving VL. The structure factor of the
$T=0$ moving VL is the same as the corresponding one of the pinned VL
[Fig.\ref{fig2}(a)]. The presence of periodic boundary conditions in our case 
prevents the occurrence of random or chaotic vortex motion  near the
critical current, as reported in early simulations with free boundary
conditions \cite{dyna0,falo}. 
In what follows we will restrict  our analysis for currents
$I<0.4$, where the collective behavior of the VL is  the dominant
physics  (at $I\sim 1$ there is a sharp increase of voltage when all
the junctions become normal, and $V\sim R_NI$ for $I\gg 1$).

\begin{figure}[t]
\begin{center}
\begin{minipage}[h]{80mm}
\centerline{\includegraphics[width=8.0cm]{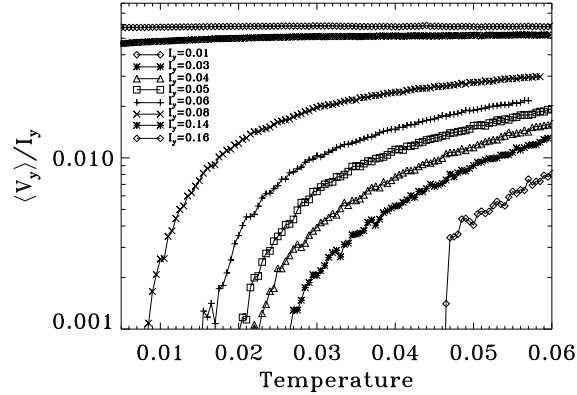}}
\end{minipage}
\caption
{$<V_y>/I_y $ vs temperature curves for different dc currents ($I_y$). 
}
\label{fig6}
\end{center}
\end{figure}

The IV curves for finite temperatures are shown in Fig.\ref{fig5}. For
temperatures below $T_M$ there is a nonlinear sharp rise in voltage 
which defines the apparent critical current $I_c(T)$. For example, we
can obtain this $I_c(T)$ with a  voltage criterion, which we choose as
$V < \frac{1}{N_t\Delta t}=10^{-4}$. In this case, we find that
$I_c(T)$ decreases with $T$, vanishing at $T_M$. It is interesting to
point out that  all the IV curves for different temperatures have a
crossing point  at $I^*=0.165$, see Fig.\ref{fig5}. A crossing in the IVs 
has also  been reported in experiments in amorphous thin
films \cite{hellerq}. For
temperatures $T>T_M$ the IV curves tend to linear resistivity for low
currents. This is shown in the inset of Fig.\ref{fig5} in a log-log plot of $R(I)=V/I$  vs
$I$, where we see that $R(I)$  tends to a low current finite value for
$T>T_M$, while it has a strong nonlinear decrease for $T<T_M$.

Let us now study the dc resistance $R=V/I$ as a function of temperature
for a given applied current in the $y$-direction (Fig.\ref{fig6}).  We start
with the perfectly ordered VL as an initial condition at $T=0$ and then
we slowly increase the temperature, keeping $I$ constant.  For currents
below the $T=0$ critical current, $I<I_c(0)$, the dc resistance is
negligibly small at low $T$, and it has a steep increase at a depinning
temperature $T_p(I)$, corresponding to the onset of vortex  motion. The
depinning temperature decreases for increasing currents, and the values
of $T_p(I)$ are coincident with the apparent critical currents $I_c(T)$
obtained from the IV curves.  For currents higher than $I_c(0)$, there
is always a large and  finite voltage for any temperature.  If 
$I_c(0)<I<I^*$, the $R(I)$ increases slightly with $T$ tending to a
constant value for  large $T$, while for $I>I^*$ the $R(I)$ decreases
with $T$.

\begin{figure}[t]
\begin{center}
\begin{minipage}[h]{80mm}
\centerline{\includegraphics[width=8.0cm]{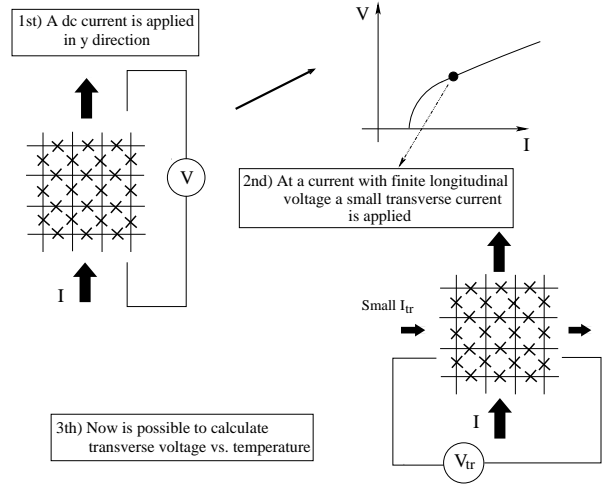}}
\end{minipage}
\caption
{How to get ``schematically'' the transverse temperature depinning.  
}
\label{fig7}
\end{center}
\end{figure}

\subsection{Transverse depinning}

What is its response to a
small current in the transverse direction when the driven vortex lattice is 
moving? Is the vortex lattice still
pinned in the transverse direction? Is there a transverse critical 
current for a moving VL?    The idea of a transverse depinning current
was introduced by Giamarchi and Le Doussal in Ref.\cite{gld} for
moving vortex systems with random pinning at  zero temperature. The
possibility of such a critical current was later questioned by Balents,
Marchetti and Radzihovsky \cite{bmr}, where it was shown that this is
not true for any  finite temperature in {\it random pinning}; however a
strong nonlinear increase of the transverse voltage was predicted at an
``effective'' transverse critical current. In the case of {\it periodic
pinning} it is  more clear that a transverse critical current will
exist at $T=0$ since it is a commensurability effect. This has been
found in the $T=0$ simulation work of Reichhardt {\it et
al} \cite{nori}.  It is also possible that this transverse critical
current will still be non-zero at $T\not=0$ in {\it periodic pinning}.
In fact, we have found in our previous work \cite{prl1} that there is a
{\it thermal} transverse depinning in a {\it periodic} system, and we
will now analyze this behavior in detail.

\begin{figure}[t]
\begin{center}
\begin{minipage}[h]{80mm}
\centerline{\includegraphics[width=8.0cm]{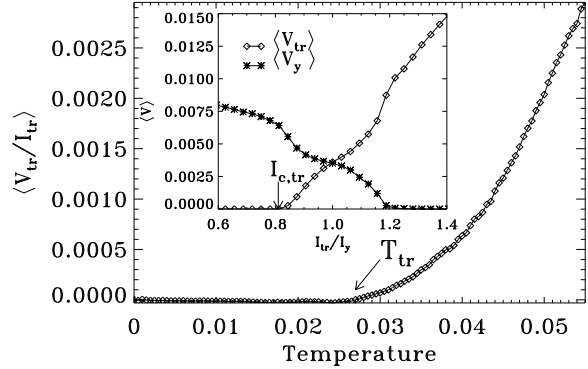}}
\end{minipage}
\caption{Transverse depinnig temperature determinations.
With a small transverse current applied, $I_x=0.01$, and
$I_y=0.16$,  $<V_{tr}/I_{tr}>$ vs $T$. Inset:   $Tr-IV$ curve
and $<V_y>$ vs $I_{tr}/I_y$  at $T=0$ and high longitudinal current,
$I_y=0.16$.     
}
\label{fig8}
\end{center}
\end{figure}

\begin{figure}[t]
\begin{center}
\begin{minipage}[h]{80mm}
\centerline{\includegraphics[width=8.0cm]{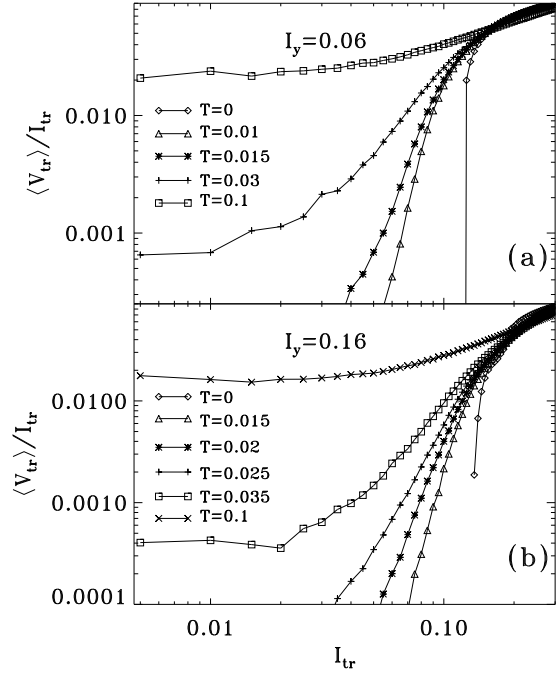}}
\end{minipage}
\caption{Transverse $IV$ curves for different temperatures:
(a) low current  $I_y < I_c(0)$, $I_y=0.06$, 
(b) high  current  $I_y > I_c(0)$, $I_y=0.16$.        
}
\label{fig9}
\end{center}
\end{figure}

First, a high longitudinal current $I>I_c(0)$ is applied at zero
temperature. Then, a current $I_{tr}$ is applied in the transverse
direction (see Fig.\ref{fig7} scheme). In this way, a transverse current-voltage characteristics
can be obtained for each $I$.  This is shown in the inset of Fig.\ref{fig8} for
$I=0.16$. We clearly see that there is a finite transverse critical
current $I_{c,tr}\approx 0.12$, which is of the order of the single
vortex pinning barrier. We also show in the inset of Fig.\ref{fig8},  how the longitudinal
voltage $V$ changes when the transverse current is varied. 
For $I_{tr}<I_{c,tr}$ the longitudinal voltage $V$ is almost constant.
At $I_{c,tr}$ there is a fast decay of $V$. When $I_{tr}=I$ we also
have $V=V_{tr}$ as expected for a drive at a degree of $\pi/4$.
Later, for $I_{tr}>I$ the vortex lattice becomes pinned in the other
direction, since now the  directions of ``longitudinal'' and
``transverse'' current are interchanged.

 Another possible measurement is to study
thermal transverse depinning. In this case, we start with a
longitudinal current  $I$, then a small transverse current is
applied,  $I_{tr}\ll I_{c,tr}$, and the temperature is slowly
increased. In this way, we can measure a transverse resistance
$R_{tr}=V_{tr}/I_{tr}$ like it was showed schematically in Fig.\ref{fig7}. 
In  Fig.\ref{fig8}  we plot this result for $I=0.16$
and $I_{tr}=0.01$. We find that for finite low temperatures $R_{tr}$ is
negligibly small within our numerical accuracy and it has a clear onset
at a  transverse critical temperature $T_{tr}$.

Let us now see how these results depend on the longitudinal current
$I$ and temperature $T$. We have calculated the transverse IV (Tr-IV)
curves  for different $I$ and $T$.  In Fig.\ref{fig9}(a) we show the Tr-IV
curves for $I=0.06$ (low current regime) and in Fig.\ref{fig9}(b) for
$I=0.16$ (high current regime). In  both cases, there is a clear
change of behavior in the Tr-IV curves when going through  a
characteristic transverse critical temperature $T_{tr}(I)$. For low
temperatures  there is a transverse critical  current  which tends to
vanish when $T$ aproachs $T_{tr}(I)$ from below.  In contrast, for
$T>T_{tr}$ there is a linear resistivity behaviour.

\begin{figure}[t]
\begin{center}
\begin{minipage}[h]{80mm}
\centerline{\includegraphics[width=8.0cm]{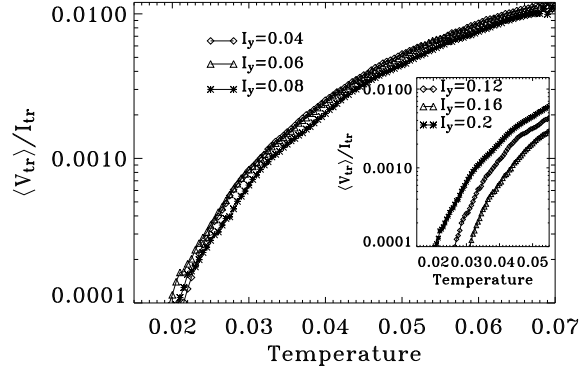}}
\end{minipage}
\caption{$V_{tr}/I_{tr}$ vs $T$ curves for different dc 
currents applied in the 
$y$ direction and small transverse current applied, $I_x=0.01$:
for low current  $I_y < I_c(0)$. Inset: for high  current $I_y > I_c(0)$.        
}
\label{fig10}
\end{center}
\end{figure}

The transverse resistivity $R_{tr}=V_{tr}/I_{tr}$ as a function of
temperature  was calculated for different longitudinal currents
(Fig.\ref{fig10}).  In all cases there is an onset of transverse response at a
given  temperature $T_{tr}(I)$. At low currents, $I<I_c(0)$, the
transverse depinning temperatures are almost constant,
$T_{tr}\sim0.02$  tending to increase slowly with $I$, see Fig.\ref{fig10}. 
On the other hand, for  $I > I_c(0)$ the transverse depinning
temperatures  increase clearly with $I$, see the inset in Fig.\ref{fig10}.

\subsection{Non-equilibrium regimes}

We will now study the different non-equilibrium regimes of vortex
driven lattices and characterize their possible dynamical transitions. 
The approach we will follow in this subsection 
is to have a fixed current applied in the system 
and vary the temperature.
In this way, we look for the possible transitions
as a function of $T$ in a similar way as was done near equilibrium in
Sec.~1.3.1.  

A few similar studies were done previously in related systems.
In Ref.~\cite{dgb} the melting transition of a moving
vortex lattice in the three dimensional XY model was studied
in this way. 
In this work a first order transition was found as a function of
temperature in a strongly driven vortex lattice. 
In Ref.~\cite{kim93} a current driven
two dimensional JJA at zero field was studied. 
The possibility of a transition  as a function of temperature
for finite  currents below the critical current was analyzed
in this case. 

In the following, we will separate our study in  three ranges of
current: low currents, $I < 0.04$ , 
intermediate currents $0.04< I < I_c(0)$ , and high currents
$I > I_c(0)$ .

\begin{figure}[t]
\begin{center}
\begin{minipage}[h]{80mm}
\centerline{\includegraphics[width=8.0cm]{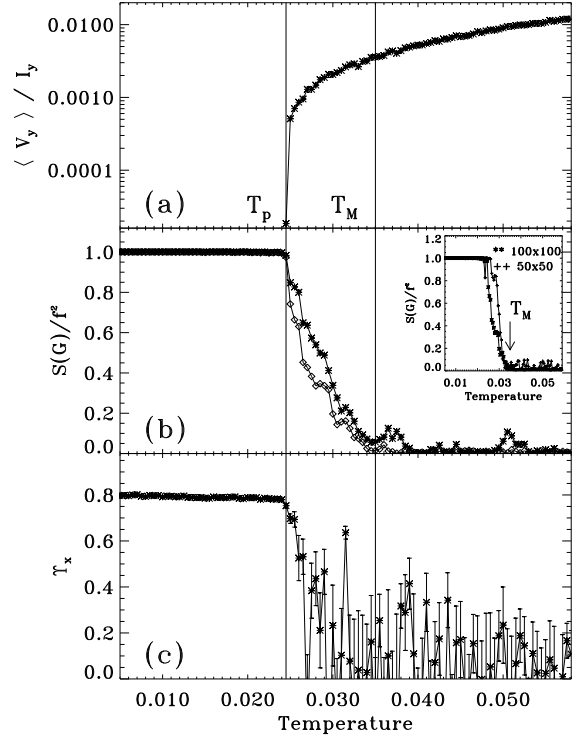}}
\end{minipage}
\caption{For low current $I< I_c(0)$, $I=0.03$:
(a) $<V_y>/I_y$ vs. $T$,
(b) $S({\bf G_1})$ $(\diamond)$ and $S({\bf G_2})$ $(\star)$ vs. $T$, inset:
 size effect in $S({\bf G_1})$, 
(c) $\Upsilon_x$ vs. $T$.
}
\label{fig11}
\end{center}
\end{figure}

\subsubsection{Low currents}

We show the results for a low current in Fig.\ref{fig11} for $I=0.03$. The
longitudinal dc response, $V/I$, is negligibly small at low $T$
and later it has a sharp increase of two orders of magnitude: this
defines the {\it depinning temperature}, $T_p$ [Fig.\ref{fig11}(a)]. At higher
temperatures, $T>T_p$, the resistance is weakly $T$-dependent. 

Below $T_p$ the vortex lattice is pinned and its structure is similar
to the ground state: a vortex lattice commensurate with the underlying
square array and tilted in the $[4a,3a]$ direction.  Above $T_p$  the
vortex lattice is moving and it has an anisotropic structural order.   
If we analyze the structure factor in two different reciprocal lattice
directions $S(G_1)$ and $S(G_2)$, 
we can see this clearly (Fig.\ref{fig11}(b)). 
Below $T_p$ the VL structure is isotropic and
$S(G_1)=S(G_2)$. Right above $T_p$ the height of the peaks decreases
with temperature, and the structure of the depinned VL is clearly
anisotropic,  $S(G_1)\not=S(G_2)$. Finally, at a melting temperature
$T_M$ the peaks vanish, and the vortex lattice melts into a vortex
liquid. In  Fig.\ref{fig11}(b) inset we compare the behavior of the Bragg peaks
for two system sizes $L=50,100$. 
We see that the $S(G_{1,2})$ are size independent below $T_p$ as it should be
expected for a pinned phase \cite{franz}. 



The helicity modulus in the
direction perpendicular to the current, $\Upsilon_x$,
decreases very slowly for $T<T_p$. Above $T_p$,
$\Upsilon_x$ has a faster decay with important
fluctuations  and tends to vanish at  $T_M$. 
For $T>T_M$, $\Upsilon_x$ oscillates around zero.
Therefore, for a small finite current the depinning and melting
transitions become separated with $T_p < T_M$.   

\begin{figure}[t]
\begin{center}
\begin{minipage}[h]{80mm}
\centerline{\includegraphics[width=8.0cm]{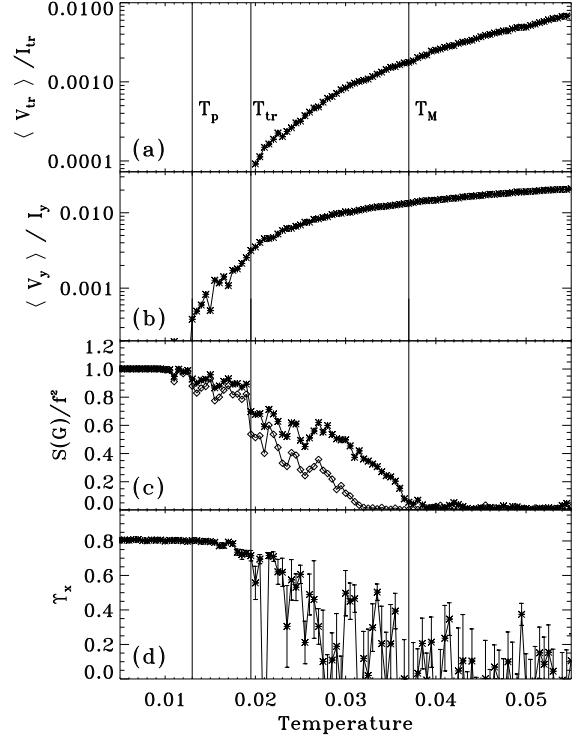}}
\end{minipage}
\caption{For $I< I_c(0)$, $I=0.06$:
(a) $<V_{tr}>/I_{tr}$ vs. $T$,
(b) $<V_y>/I_y$ vs. $T$,
(c) $S({\bf G_1})$ $(\diamond)$ and $S({\bf G_2})$ $(\star)$ vs. $T$,
(d) $\Upsilon_x$ vs. $T$.
}
\label{fig12}
\end{center}
\end{figure}
\begin{figure}[t]
\begin{center}
\begin{minipage}[h]{80mm}
\centerline{\includegraphics[width=8.0cm]{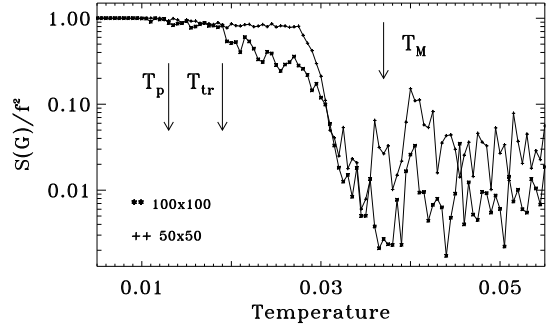}}
\end{minipage}
\caption{For intermediate current $I< I_c(0)$, $I=0.06$: size effect in $S({\bf G_1})$.
}
\label{fig13}
\end{center}
\end{figure}
\subsubsection{Intermediate currents}

At intermediate currents, $0.04<I<I_c(0)$, a {\it new} transition
appears: the {\it transverse depinning} of the  moving vortex lattice.
As discussed in Sec.~1.3.3, one can measure transverse depinning by applying
a small transverse current while the VL is driven with a fixed 
longitudinal current.
This is shown in Fig. \ref{fig12}(a) for the case 
of $I=0.06$ and a small transverse
current, $I_{tr}=0.01$. We see that
there is an onset of transverse voltage  at $T_{tr}=0.019$.
We can also see that this transverse depinning temperature is
above the depinning temperature $T_p$ for longitudinal
resistance [Fig.\ref{fig12}(b), $T_p=0.013$], and below 
the melting temperature $T_M$ for the vanishing of the Bragg
peaks [Fig.\ref{fig12}(c), $T_M=0.037$]. 
Therefore, this transition occurs at an intermediate
temperature between the depinning
and the melting transitions, $T_p<T_{tr}<T_M$.
We also show in Fig.\ref{fig12}(d) that 
the helicity modulus  begins to fall down slowly 
at $T_{tr}$, while for $T_{tr}<T<T_M$ it has strong fluctuations,
being difficult to interpret its behavior in this case.

One can see that the intensity of the Bragg
peaks has a greater dependence with size for 
$T_{tr}<T<T_M$ when compared with the $T_p<T<T_{tr}$ regime in  Fig.\ref{fig13}.
For large temperatures $T>T_M$, in the liquid phase, the value
of $S(G_{1,2})$ in general is strongly size dependent, 
since it should go as $L^{-2}$ \cite{franz}.
On the other hand,
for $T_p<T<T_M$ we find that the intensity of the Bragg
peaks is weakly dependent on system size.  
The clear change of behavior of the size dependence gives a 
good criterion to determine  $T_M$ (Fig.\ref{fig13}).

\begin{figure}[here]
\begin{center}
\begin{minipage}[h]{120mm}
\centerline{\includegraphics[width=12.0cm]{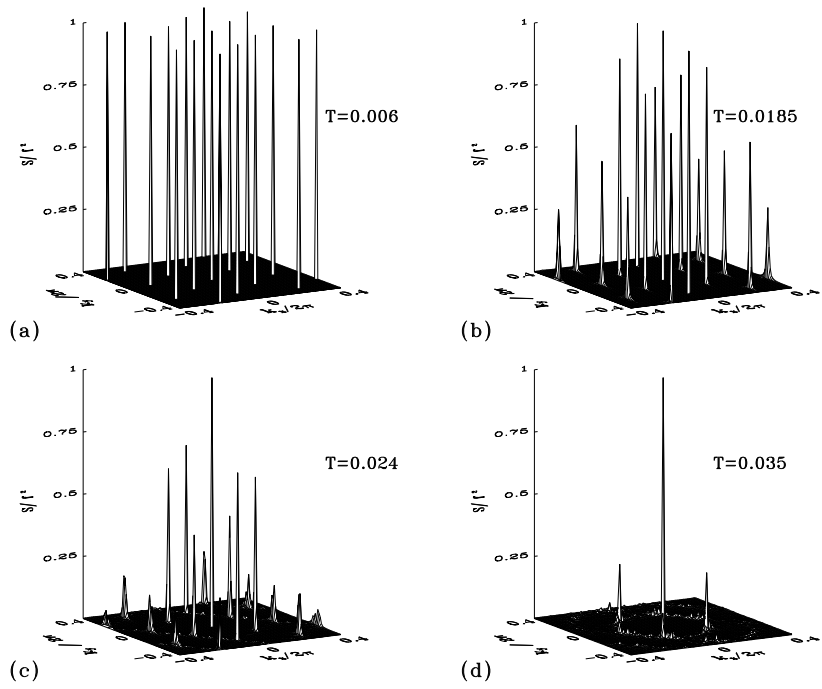}}
\end{minipage}
\caption{Intensity plot of the structure factor $S({\bf G})$ at $I<I_c(0)$,
 $I=0.06$ for different temperatures:
(a) $T < T_p$, $T=0.006$,
(b) $T_p < T < T_{tr}$, $T=0.0185$,
(c) $T_{tr} < T < T_M$, $T=0.024$,
(d) $T \leq T_M$, $T=0.035$.         
}
\label{fig14}
\end{center}
\end{figure}

It is interesting to study in more detail the behavior of 
the structure factor through all these transitions.
In Fig.\ref{fig14} we show examples of the structure factor $S({\bf k})$ 
at temperatures in the different regimes.
In the pinned phase, the $S({\bf k})$ is nearly the same as in
the ground state with delta-like Bragg peaks, see Fig.\ref{fig14}(a).
For the moving VL, we can see that 
there is 
less anisotropy  in the {\it transversely pinned}
regime $T_p<T<T_{tr}$  [Fig.\ref{fig14}(b)]
than in the {\it floating} regime $T_{tr}<T<T_M$ [Fig.\ref{fig14}(c-d)].
 Moreover, near $T_M$ the VL structure becomes  strongly anisotropic,
with the peak at ${\bf G}_1$ much larger than the peak at ${\bf G}_2$ (see  Fig.\ref{fig14}(d) and Fig.\ref{fig13}).

The anisotropy of the Bragg peaks of the moving VL  (in the regimes at
$T_p<T<T_M$) has two  characteristics: (i) the {\it width} of the peaks
increases  with $T$ in the direction of the applied current (the direction
perpendicular to the vortex motion), and (ii)  the {\it height} of the peaks
decreases in the direction of vortex motion. This can be observed in the
sequence of structure factors shown in Fig.\ref{fig14}. 

\clearpage
\begin{figure}[t]
\begin{center}
\begin{minipage}[h]{80mm}
\centerline{\includegraphics[width=8.0cm]{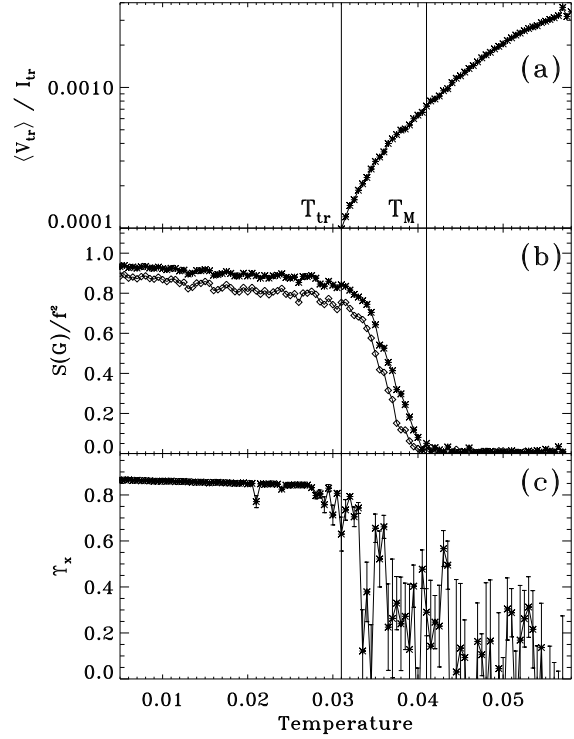}}
\end{minipage}
\caption{ For high current $I> I_c(0)$, $I=0.16$:
(a) $<V_{tr}>/I_{tr}$ vs. $T$,
(b) $S({\bf G_1})$ $(\diamond)$ and $S({\bf G_2})$ $(\star)$ vs. $T$,
(c) $\Upsilon_x$ vs. $T$.
}
\label{fig15}
\end{center}
\end{figure}

\subsubsection{High currents}

\begin{figure}[t]
\begin{center}
\begin{minipage}[h]{80mm}
\centerline{\includegraphics[width=8.0cm]{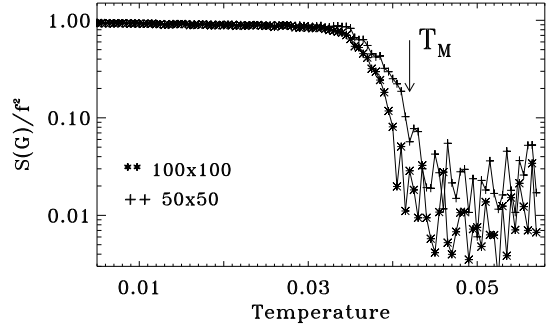}}
\end{minipage}
\caption{ For high current $I> I_c(0)$, $I=0.16$:
 size effect in $S({\bf G_2})$  vs. $T$.
}
\label{fig16}
\end{center}
\end{figure}

In the case of  currents  larger than $I_c(0)$, the vortex lattice
is already depinned at $T=0$. As we have seen in Sec.~1.3.3, this
zero-temperature moving VL has a finite transverse critical current,
and therefore it is pinned in the transverse direction.
When we slowly increase temperature from this state, we find
that the transverse resistive response is negligible for 
finite low temperatures. At a  temperature $T_{tr}$ there is a
jump to a finite transverse resistance $R_{tr}=V_{tr}/I_{tr}$.
For example,  this is shown for  $I=0.16$ with a small transverse current,
$I_x=0.01$ in Fig.\ref{fig15}(a). 
The vortex lattice has an anisotropic structural order for all temperatures,
{\it i.e}, $S({\bf G_1})\neq S({\bf G_2})$,
and the height of the Bragg peaks vanishes at $T_M$, Fig.\ref{fig15}(b). 
The transverse helicity modulus $({\Upsilon_x})$  is 
almost constant for low
temperatures and starts to decrease at $T_{tr}$ presenting strong fluctuations 
for $T>T_{tr}$ , Fig\ref{fig15}(c). 
In Fig.\ref{fig16} we show $S({\bf G_2})$ for different sizes, $L=50,100$, 
and we see that $T_M$ is size independent. 
Similar behaviour is found for $S({\bf G_1})$.

The analysis of the heights of the peaks in $S({\bf k})$ as a function of
system size $L$ is a good indicator of the translational correlations in the
system. This dependence is well known for two dimensional lattices 
\cite{franz}. For a pinned solid $S({\bf G})\sim1$, 
for a floating solid, $S({\bf G})\sim
L^{-\eta_G(I,T)}$ with $0<{\eta_G(I,T)}<2$, being this dependence a signature
of quasi-long range order, 
and for a normal liquid $S({\bf G})\sim L^{-2}$.  To
assure the existence of algebraic translational 
\begin{figure}[t]
\begin{center}
\begin{minipage}[h]{80mm}
\centerline{\includegraphics[width=8.0cm]{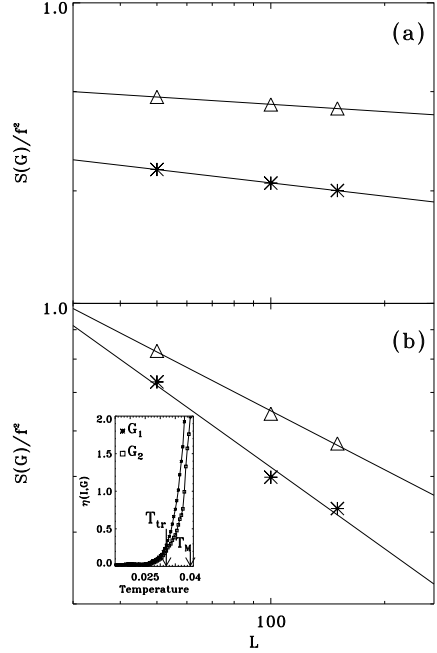}}
\end{minipage}
\caption{ Finite size
analysis and power law fit of $S({\bf G})\sim L^{-\eta_G}(I,T)$ for a moving
vortex lattice at $I>I_c(0)$, $I=0.16.$ We obtain: (a) for  $T < T_{tr}$,
$T=0.02$,  $\eta_{G_1}=0.023$ $(\star)$ and $\eta_{G_2}=0.013$, $(\triangle)$,
(b) for $T_{tr} < T < T_M$, $T=0.035$, $\eta_{G_1}=0.471$ $(\star)$ and 
$\eta_{G_2}=0.34$ $(\triangle).$  Inset: at same $I$, $\eta(I,{\bf G})$ vs $T$, 
for ${\bf G_1}$and ${\bf G_2}$. 
}
\label{fig17}
\end{center}
\end{figure}
correlations, 
we have done 
this scaling study for currents $0.02<I<0.2$ and different temperatures
for system sizes of $L=50,100,150$. 
For the cases corresponding to the pinned regime ($T<T_p$)
we found $\eta_G\approx0$, as expected. 
In the transversely pinned regime, we show a case in Fig.\ref{fig17}(a),
we find a power law fitting with  very small values of ${\eta_G}(I,T)$.
In the floating regime, we find larger values of $\eta_G$,
we show a case in Fig.\ref{fig17}(b).
In all the cases we have obtained that 
 $0<{\eta_G}(I,T)<2$ for $T_p<T<T_M$. 
Therefore, this finite size analysis shows the existence of 
quasi-long range order in the moving VL.
Also, we find that ${\eta_{G_1}} > {\eta_{G_2}}$ for all currents
and temperatures.   
The power-law exponent ${\eta_G}$ can be studied
at a constant current,  as a function of temperature. 
This is shown in the inset of Fig.\ref{fig17}(b), for $I=0.16$
and different reciprocal lattice vectors. The exponent ${\eta}$ is
finite for the complete temperature range.
For $T<T_{tr}$ it has a small value $\eta_G\approx 0.01$. 
It has a fast increase near $T_{tr}$, 
where there is also a clear difference between ${\eta_{G_1}}$
and ${\eta_{G_2}}$.
Finally, it reaches a  value of  ${\eta=2}$ near $T=T_{M}$.

\subsection{Orientational pinning effects}

A very interesting  characterization of the different regimes showed in the precedent subsection
can
be obtained by  studying  the effects of varying current direction
\cite{prbnos,stroud99,choi}. 
We apply a current $I$ at an angle $\phi$ with respect to the $[10]$
lattice direction, $I_x=I\cos\phi,  I_y=I\sin\phi$ (see Sec.~1.2.3).
We study the voltage response when varying the orientation $\phi$ of the drive
while keeping fixed the amplitude $I$ of the current.  In the parametric
curves of $V_{y}(\phi )$ vs. $V_{x}(\phi )$ can we analyze the
breaking of rotational symmetry  in the different regimes of $I$ and $T$. 
In the case of rotational symmetry this kind of plot should
give a perfect circle. However, 
the square symmetry of the Josephson lattice 
will show up  in the shape of the curves. In what follows 
  we start studying  the motion of single vortices in JJA (1), to show later 
the breaking of rotational invariance in diluted vortex lattices (2) and how this effect could be used to characterized the 
  dynamical regimes described previously.   The first case of a single vortex is studied before for its simplicity.  In order to understand the phenomenon
 of breaking of rotational symmetry  is better to begin without including the collective effects of vortex lattices. In addition, experimental 
evidences were found simultaneously for Dr. H. Pastoriza and collaborators \cite{prbnos} for this case. And very nice agreement between numerical
 and experimental results was found.

\subsubsection{Breaking of rotational invariance in JJA with a single vortex}
\begin{figure}[t]
\begin{center}
\begin{minipage}[h]{70mm}
\centerline{\includegraphics[width=7.0cm]{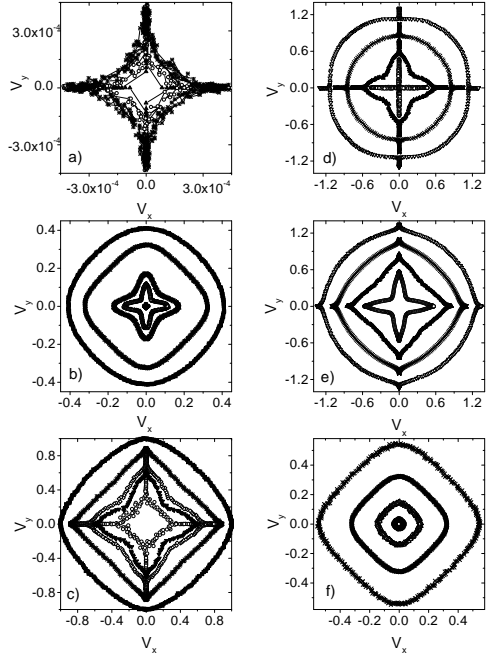}}
\end{minipage}
\caption{Simulation results of the
parametric curves $V_y(\phi)$ vs. $V_x(\phi)$.
(a) $I=0.2$, for (going
outwards from the center) $T=0$, $T=0.01$ and $T=0.05$; 
(b) $I=0.6$ for $T=0.3$, $T=0.5$, $T=0.6$, $T=1.0$ and $T=1.4$;
(c) $I=1.2$ for $T=0$, $T=0.05$, $T=0.1$, $T=0.3$ and $T=1.0$. 
(d) $T=0.05$  for (going
outwards from the center) $I=1.0$, $I=1.2$, $I=1.4$ and $I=1.6$;
(e) $T=0.2$ for $I=1.0$, $I=1.2$, $I=1.4$ and $I=1.6$; 
(f) $T=1.0$ for $I=0.2$, $I=0.4$, $I=0.6$ and $I=0.8$.
}
\label{fig18}
\end{center}
\end{figure}

The square lattice has two directions of maximum symmetry: the [10] and the
[11] directions (and the ones obtained from them by $\pi/2$ rotations),
which correspond to the directions of reflection symmetry. When the current
bias is in the [10] direction, the angle of the current is $\phi=0$, and we
call it a ``parallel" bias. When the current bias is in the [11] direction,
the angle of the current is $\phi=\pi/4=45^o$, and we call it a ``diagonal"
bias (see squeme in Fig. \ref{fig1}).

In the case of the parallel bias we find  that the transverse voltage is zero
(in agreement with the reflection symmetry), see details in Ref.~\cite{prbnos}. 
This corresponds to vortex
motion in the direction perpendicular to the current ($\theta_t=0$). A perfect agreement with experimental results was found \cite{prbnos}.

In the IV curve for the longitudinal voltage we find numerically a
critical current corresponding to the single vortex depinning $I_c^{[10]}=0.1
$ \cite{jjpin, prbnos}. In the case of the perfect diagonal bias, which is only attainable in the 
numerical model, we obtain similar results as in the
parallel bias case. The transverse voltage is zero and therefore the vortex
moves perpendicular to the current ($\theta_t=0$). In agreement with the prediction of Halsey~\cite{halsey}, the numerical 
IV curve for $V_l$
has a critical current of $I_c^{[11]}=\sqrt{2}I_c^{[10]}=0.1414$.   The onset
of the resistive regime is also multiplied by a factor of $\sqrt{2}$.

For orientations different than the symmetry directions, we always find 
 a finite transverse voltage (as well as was found in experiments \cite{prbnos}). 
In order to see this, we  study the voltage
response when varying the orientation $\phi$ of the drive while keeping
fixed the amplitude $I$ of the current.  If we analyze  the transverse
angle $\theta _{t}=\arctan ({V_{t}}/{V_{l}})$ as a function of the angle of
the current $\phi$  we observe that  $\theta _{t}$ vanishes only in the maximum
symmetry directions corresponding to angles $\phi =0,\pm 45^{o},\pm
90^{o},\dots$. Furthermore, we see that for
orientations near $\phi=0$, the transverse angle basically follows the
current angle: $\theta _{t}\approx -\phi $. This is an indication that
vortex motion is pinned in the lattice direction [10], since $V_{y}\approx 0$,
 meaning that the voltage angle is $\theta _{v}\approx 0$. 
Whenever the voltage response is
insensitive to small changes in the orientation of the current, we 
will call this phenomenon {\it orientational pinning}. On the other
hand, near $\phi=45^{o}$ the transverse angle changes rapidly. 

A more direct evidence of the breaking of rotational symmetry 
can be seen in the parametric curves of $V_{y}(\phi )$
vs. $V_{x}(\phi )$. In Fig.~\ref{fig18} we plot the values obtained 
for the voltages $V_{y}$ and 
$V_{x}$ when varying the orientational angle $\phi $ for different values of
the current amplitude $I$ and the temperature. In the case of rotational
symmetry we should have a perfect circle. In the set of plots of Fig.~\ref{fig18}(a-c),
the current amplitude is fixed and the temperature is varied. In Fig.~\ref{fig18}(a) we
have $I=0.2$, near the onset of single vortex motion in the regime.
In this case most of the points are either on the axis $V_{x}=0$ or on the
axis $V_{y}=0$, indicating strong orientational pinning in the lattice
directions [10] or [01]. When increasing $T$ the orientational pinning
decreases and the length of the ``horns'' in the $x$ and $y$ axis decreases.
Fig.~\ref{fig18}(b) corresponds to $I=0.6$ , when the
vortex is moving fast. In this case the horns have disappeared and
orientational pinning is lost. However, the breaking of rotational symmetry
is still present in the star-shaped curves that we find at low $T$. The dip
at $45^0$ in the stars are because in this direction the voltages are
minimum, since the critical current is maximum in this case, $I_c^{[11]}=0.1\sqrt{2}$.
 When increasing the temperature, the stars tend to the circular
shape of rotational invariance. Above the onset of the resistive regime 
the ``horned'' curves reappear. 
In this case the orientational pinning
corresponds to the locking of ohmic dissipation in the junctions in one of
the lattice directions, either [10] or [01]. Once again, when increasing $T$ the
horned structure shrinks, and the curves evolve continuously from square shapes
to circular shapes.

The variation with current of the rotational parametric curves for a fixed
temperature is shown in the numerical results of 
Figs.~\ref{fig18}(d-f). At a low temperature, $T=0.05$ we
clearly see the horned structure of the curves for almost all the currents
and even for large currents the circular curves have ``horns'', see
Fig.~\ref{fig18}(d). At an intermediate temperature, $T=0.2$, there are still some
signatures of the orientational pinning [Fig.~\ref{fig18}(e)], while for $T=1.2$ all the curves
are smooth and rounded with a slightly square shape [Fig.~\ref{fig18}(f)].\\

{\it Transverse voltage near the [1,1] direction}\\

\begin{figure}[t]
\begin{center}
\begin{minipage}[h]{70mm}
\centerline{\includegraphics[width=7.0cm]{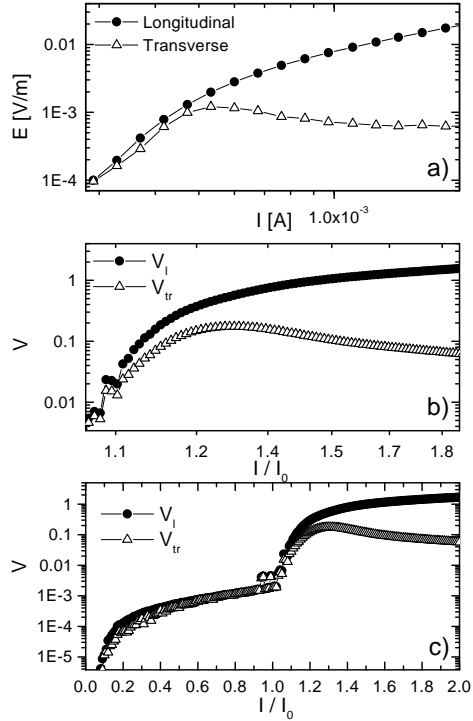}}
\end{minipage}
\caption{(a) Experimental results of longitudinal and transverse voltage for
a current near the [11] direction, $\phi=45^o\pm5$ at $T=1.25$ K. (b)
Longitudinal and transverse voltage obtained numerically for $\phi=40^o$ at $%
T=0.02\, \hbar I_0/2ek_B$. (c) Idem (b) for an extended current range.}
\label{fig19}
\end{center}
\end{figure}
In Fig.~\ref{fig19}(a) we show  experimental voltage-current characteristics for an
array of $100\times 1000$ junctions at a low temperature $T=1.25K$ and at a
low magnetic field (results from Dr. Hernan Pastoriza's Group, Argentina).
 The current is applied nominally in the [11] direction,
but a small misalignment is possible in the setup of electrical contacts,
therefore $\phi =45^{o}\pm 5^{o}$. We see that for low currents there is a
very large value of the transverse voltage $V_{t}$, which is nearly of the same
magnitude as the longitudinal voltage $V_{l}$. The transverse voltage is
maximum at a characteristic current $I_{m}$. Above $I_{m}$, $V_{t}$
decreases with increasing current while $V_{l}$ increases. It is remarkable
that these results are very different from the IV curve obtained numerically,
 where $
V_{t}=0$ at $\phi =45^{o}$. However, if we assume a misalignment of a few
degrees with respect to the [11] direction we can reproduce the experimental
results. In Fig.~\ref{fig19}(b) we show the IV curves obtained numerically for $\phi
=40^{o}$ and $T=0.02$. We see that for low currents $V_{t}$ is close to 
$V_{l}$: $V_{t}\lesssim V_{l}$, similar to the experiment, and later $V_{t}$
has a maximum at a current $I_{m}\approx 1/\cos \phi \approx 1.3$. This
corresponds to the current for which the junctions in the $x$-direction
become critical ($I_{x}=1$). 
The highest transverse voltage can be obtained for
orientations near $\phi=45^o$ as we explained before.

Therefore a slight misalignment of the array
from the [11] direction is useful for studying both experimentally and
numerically the behavior of the transverse voltage as a function of current
and temperature.\\

{\it Transverse voltage near the [1,1] direction in disordered JJA}\\

Before we  studied the breaking of rotational invariance in  square JJA\cite{prbnos} at finite
temperatures analyzing  transport properties and we obtained that the transverse
voltage vanishes only in the directions of maximum symmetry of the square lattice: the [10] and
[01] direction (parallel bias) and the [11] direction (diagonal bias).
For diagonal bias  this result is highly unstable against small variations of the angle of 
the applied current,  leading
to a rapid change from zero transverse voltage to a large transverse voltage within a few degrees.
Now we will show  that a small amount of disorder induces finite transverse
voltage  in square JJA with a {\it perfect diagonal bias}. The transverse voltage as a function of
current presents a  peak which does not depend strongly with disorder
for moderate disorder strength.
This result is experimentally relevant since  samples always could present a  small amount of
disorder  due to tiny fabrication defects in addition to a possible misalignment in the setup of
electrical contacts. 
\begin{figure}[t]
\begin{center}
\begin{minipage}[h]{80mm}
\centerline{\includegraphics[width=8.0cm]{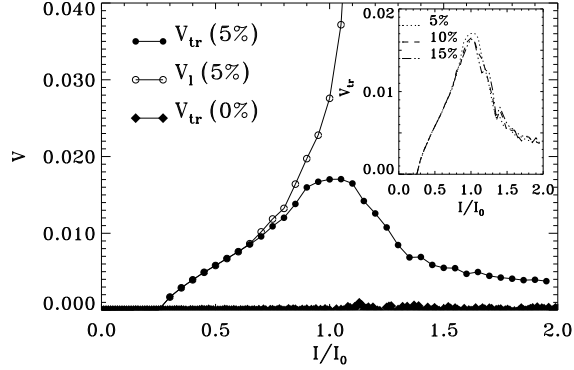}}
\end{minipage}
\caption{Numerical IV characteristics  in  a system of $32\times32$ junctions with
 diagonal bias, at $T=0.02$,and $5$\% of disorder compared with transverse voltage   without disorder.
 Inset: transverse voltage
peak for $5$\%, $10$\%  and $15$\% of disorder.
}  
\label{fig20}
\end{center}
\end{figure}

We simulate an square
 JJA with disorder in the  critical currents using the RSJ model introduced before
 in Sec.~1.2.1.
The modification to the current flowing in the junction between two superconducting
islands is the following
\begin{equation}
I_{\mu}({\bf n})= g_{\mu}({\bf n}) I_0 \sin\theta_{\mu}({\bf n}) + 
                  \frac{\Phi_0}{2\pi c R_N} 
		  \frac{\partial \theta_{\mu}({\bf n})}{\partial t}
		  +\eta_{\mu}({\bf n},t),
\label{rsjd}
\end{equation}
where   disorder is introduced
through the factor   $g_{\mu}({\bf n})=1+ \Delta (\xi_\mu({\bf n}) - 1/2)$. $\xi$ is a uniform
distributed random variable in $[0,1]$. Therefore 
$I_{0}(1 - \frac{\Delta}{2})  \leq g_{\mu}({\bf n})I_{0} \leq 
I_{0} (1 +  \frac{\Delta}{2})$. 
Again we
calculate the time average of the total voltage in both directions, $V_x$ and $V_y$, and transverse and logitudinal voltages from them, as in Eq.~\ref{tv} and \ref{lv}.

\begin{figure}[t]
\begin{center}
\begin{minipage}[h]{80mm}
\centerline{\includegraphics[width=8.0cm]{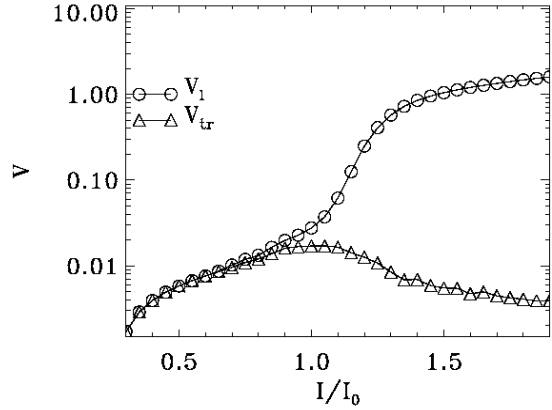}}
\end{minipage}
\caption{Numerical simulations: IV curves for  $\Delta=0.1$
($V_l (\circ)$ and $V_{tr} (\triangle)$).
}  

\label{fig21}
\end{center}
\end{figure}

In Fig.\ref{fig20} we show  current-voltage characteristics, at $T=0.02$ (in units of the Josephson
energy, $k_BE_J$) obtained numerically for an array of $32\times32$
 junctions  without ($\Delta = 0$) and with disorder($\Delta = 0.1$, {\it i.e.}
 $\pm 5$\% variation)  in the critical currents. The
 dc current is applied in the diagonal direction ($\phi=45^o$), $[1,1]$.  We    clearly see the
 appearance of finite transverse voltage, $V_ {tr} \neq 0$, with a $5$\%
  of disorder  in
  the critical currents. For low currents longitudinal voltage is equal to 
  transverse voltage  and
   they start to separate when the vortex-antivortex
  pairs density  increases considerably  upon increasing current. 
  $V_{tr}$ presents a maximum around $I=I_0$ and for 
  larger currents tends to zero. This behaviour is in contrast 
  to the one obtained before without disorder, where $V_{tr}=0$  for all currents. 
  We calculate IV characteristics as a
 function of the intensity of disorder and we observe, as we show in the inset, 
 that the transverse
 voltage  is almost independent of  $\Delta$ in the range considered,
 $0.1 \le \Delta \le 0.6$.

In Fig.\ref{fig21} we show in a log-log scale our previous results   in order to compair them with the experimental results,
 obtained in 
an array of $100\times 1000$ proximity--effect Pb/Cu/Pb junctions, at a low temperature $T=1.25K$ and at a
low magnetic field (see Fig.~\ref{fig19}). 
Qualitatively  we get a good agreement between experiments and simulations
 now in the case of diagonal bias, in opposite to the results obtained without disorder and diagonal bias.

 The simulation results presented up to this point were
focused in the motion of a single vortex in the periodic pinning of a square
JJA. The vortex collective effects,  for fields $f>1/L^2$, will be discussed below.

\begin{figure}[t]
\begin{center}
\begin{minipage}[h]{90mm}
\centerline{\includegraphics[width=9.0cm]{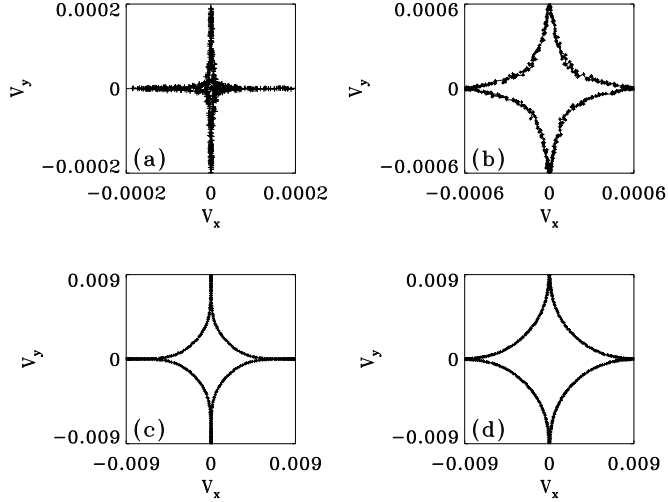}}
\end{minipage}
\caption{ Parametric curves $V_y(\phi)$ vs. $V_x(\phi)$ for different dc current
applied and temperatures:
(a) $|I_y|=0.06$ and $T=0.015$,
(b) $|I_y|=0.06$ and $T=0.03$,
(c) $|I_y|=0.16$ and $T=0.015$,
(d) $|I_y|=0.16$ and $T=0.035$.        
}
\label{fig22}
\end{center}
\end{figure}

\subsubsection{Breaking of rotational invariance in diluted vortex lattices}

In the previous subsection we showed that 
the ``diagonal''
[11] direction is unstable against small changes in the angle
$\phi$, while the [10] and [01] directions are the 
preferred directions for single vortex motion \cite{prbnos}.
This shows as ``horns'' in  parametric $V_y$ vs. $V_x$  plots,
which are finite segments of points lying in the $x$ or the 
$y$ axis. This implies the existence of transverse 
pinning in these directions 
(thus, it corresponds to orientational pinning).
Now we return to the case of diluted  vortex lattices   ($f=1/25$) and we perform  the same  analysis as before for the different regimes
of the  moving VL obtained in Sec.~1.3.4.

In Fig.\ref{fig22} we plot the voltages $V_{y}$ and  $V_{x}$ 
when varying the orientational angle
$\phi $ for different  current amplitudes $I$ 
and temperatures $T$.
In Fig.\ref{fig22}(a) we have $I=0.06$ and $T=0.015$,  corresponding to the regime of  
a transversely pinned lattice. 
In this case most of the points are lying either on
the axis $V_{x}=0$ or on the axis $V_{y}=0$, 
indicating strong orientational pinning in the symmetry lattice
directions [10] and [01]. 
When increasing $T$ the orientational pinning
decreases and the length of the ``horns'' in the $x$ and $y$ axis decreases.
Fig.\ref{fig22}(b) shows results for $I=0.06$ and $T=0.03$, which correspond
to $T>T_{tr}$ when there is a finite
transverse resistance. In this case the horns have disappeared
and orientational pinning is lost. 
However, the breaking of rotational symmetry
is still present in the star-shaped curves. 
Also in the high current regime, Fig.\ref{fig22}(c), for a low temperature  $T=0.015$
($T<T_{tr}$) we see that there is orientational pinning with the presence of
horns, which again disappears for $T>T_{tr}$ as it is shown in Fig.\ref{fig22}(d) at
$T=0.035$. 
Finally for $T\gg T_M$, deep inside
in the liquid phase,  the stars tend to the circular shape 
of rotational invariance.

Therefore we have shown that orientational pinning  
is a useful phenomenon to characterize the different dynamical regimes in JJA,
 both numerically and experimentally.

\subsection{Summary and discussion} 

\begin{figure}[t]
\begin{center}
\begin{minipage}[h]{80mm}
\centerline{\includegraphics[width=8.0cm]{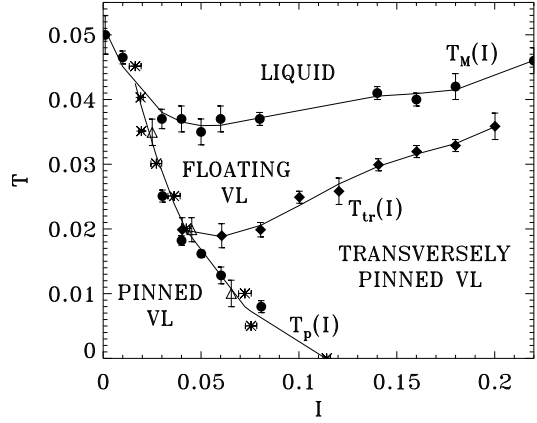}}
\end{minipage}
\caption{$I-T$ Phase diagram for $f=1/25$. $T_M(I)$ line is obtained from
$S({\bf G})$ vs. $T$ curves $(\bullet)$. $T_p(I)$ line is obtained from IV
curves $(\star)$, from $S({\bf G})$ vs. $T$ curves $(\bullet)$ and from
$\langle V_y\rangle$
vs. $T$ curves $(\triangle)$. $T_{tr}(I)$ curve is obtained from
$\langle V_{tr}\rangle$ vs.
$T$ curves $(\blacklozenge)$. 
}
\label{fig23a}
\end{center}
\end{figure}

With all the information of the
previous sections we obtain the current-temperature phase diagram
shown in  Fig.\ref{fig23a}.
For finite currents we have been able to identify
three different regimes,  a {\it pinned VL} for $T<T_p(I)$,  
a {\it transversely pinned VL} for $T_p(I)<
T < T_{tr}(I)$, and a {\it floating VL} for $T_{tr}(I) < T <
T_{M}(I)$. 
It is, however, difficult to define if the temperatures
$T_p, T_{tr}, T_M$ correspond to either
phase transitions or to dynamical crossovers.  
The comparisons we have made of the behavior of voltages (longitudinal and
transversal), structure factor and helicity modulus   show that
 something is happening at these temperatures.
Also, the
comparison for different system sizes of the behavior
of $S({\bf G})$ suggest  transitions for $T_p, T_{tr}, T_M$.
The transverse helicity modulus has strong fluctuations for 
$T_{tr}< T < T_{M}$. These  fluctuations are not reduced when
increasing the simulation time in a factor of $10$. This
could mean that actually the range of temperatures 
of $T_{tr}< T < T_{M}$ corresponds to a long crossover region towards
a liquid state.  Also, one may question if  the use of the helicity
modulus for these far from equilibrium states is correct, since $\Upsilon_x$
has been defined from the response of the equilibrium  free energy
to a twist in the boundary condition.
Most likely, the large fluctuations in $T_{tr}< T < T_{M}$ are due to
the very unstable and history-dependent steady states we find in this regime. 
For example the steady states obtained when increasing temperature from $T=0$
at fixed current differ from the steady states obtained when increasing
current at fixed temperature in this regime. They differ in the degree
of anisotropy and intensity of the Bragg peaks in the structure factor.
We think that this reflects the fact that the VL is unpinned in all
directions and the orientation of the moving state will depend on the
history for this regime. 

In any case, we can clearly  distinguish different dynamical  states
which define the regimes shown in the phase diagram of Fig.\ref{fig23a}. From the
experimental point of view these transitions (or crossovers) are
possible to measure. The depinning temperature $T_p$ can be obtained
from resistance measurements at finite currents. The transverse 
current-voltage characteristics and the transverse resistivity can be
measured for different longitudinal currents and temperatures, and therefore
 $T_{tr}$ could be obtained. The helicity modulus can be measured with the
two-coil technique \cite{jjarev}. It could be very interesting to see
the results of such measurements at finite currents. 
The structure factor and melting transition can not be measured
directly. However, in the presence of an external rf current, 
the disappearance of Shapiro steps can indicate the melting
transition \cite{harris}, since they are
sensitive to the translational order of  the vortex lattice \cite{shapiro}.
Therefore, we expect that the results obtained here could motivate
new experiments for fractional or submatching
fields in Josephson junction arrays as well as in other
superconductors with periodic pinning.

After our articles (see Ref.\cite{prl1},\cite{prbnos},\cite{prb2}) several experimental works appeared.
First of all, D. Shal\'om and H. Pastoriza designed a new two coil kinetic inductance technique in order to measure transport
 anisotropy  induced by the applied current itself. They found for first time experimental evidence for the anisotropic character of the current driven 
diluted vortex lattices states in JJA   \cite{diego1}. Later they improved the previous technique, implementing rectangular coils with high aspect radio 
which are now lithographically fabricated on top of the sample, separated for isolation layers \cite{diego2}.  

Regarding  numerical works, it is very interesting to mention that recently the disappearance of Shapiro steps was used to  indicate the melting
transition for another submatching field, $f=1/2$ \cite{kim04}, system which we are going to discuss in next section.

About our results on orientational pinning we would like to do a series of remarks and comments.
 In the egg-carton potential of a square JJA  there are pinning barriers for
vortex motion in all the directions. The direction with the lowest pinning
barrier is the [10] direction. Therefore the strong orientational pinning we
find here is in the direction of the lowest pinning for motion, i.e. the
direction of easy flow for vortices. The presence of a strong orientational
pinning leads to a large transverse voltage when the systems is driven away
from the favorable direction, to the existence of a critical angle and to a
transverse critical current. On the other hand, the [11] direction is the
direction of the largest barrier for vortex motion in the egg-carton
potential. In this case, the behavior is highly unstable against small
variations in the angle of the drive, leading to a rapid change from zero
transverse voltage to a large transverse voltage within a few degrees.
Any misalignment of the current/voltage contacts as well as disorder
in the junctions critical currents \cite{jlt} can lead to a large
transverse voltage for arrays with a diagonal bias.
This explains the transverse voltage observed experimentally in
JJA\ driven near the diagonal [11] orientation \cite{prbnos}.

An analogous effect of orientational pinning has also been seen in
experiments on YBCO
superconductors with twin boundaries \cite{twins}. In this case, due to the
correlated nature of the disorder, the direction for easy flow is the
direction of the twins. A similar effect of horns in the parametric voltage
curves are therefore observed in the direction corresponding to the twins.
Also transport measurements when the sample is driven at an angle with
respect to the twin show a large transverse voltage.

It is interesting to compare with the angle-dependent transverse voltage
calculated for d-wave superconductors \cite{vicente}. Also in this case, the
transverse voltage vanishes only in the [10] and [11] directions.  However
the $\theta_{t}$ vs. $\phi $ curves are smooth in this case, since
$\tan\theta_t\propto \sin4\phi$ \cite{vicente}. This is because there is no
pinning and the transverse voltage is caused only by the intrinsic nature of
the d-wave ground state. On the other hand, the breaking of rotational
symmetry studied here is induced by the pinning potential, and it results in
non-smooth responses like ``horned'' parametric voltage curves, critical
angles, transverse critical currents, etc.

In superconductors with a square array of pinning centers, typically the
pins are of circular shape and the size of the pins is much smaller than the
distance between pinning sites \cite{nori0,nori,nori2}. In this case, the
pinning barriers that vortices find for motion are the same in many
directions. Therefore it is possible to have orientational pinning in many
of the square lattice symmetry directions. This explains the rich structure
of a Devil's staircase observed recently in the simulations of Reichhardt
and Nori, where each plateau corresponds to orientational pinning in each of
the several possible directions for orientational pinning.  This interesting
behavior is not possible in JJA, however, since the egg-carton pinning
potential corresponds to the situation of square-shaped pinning centers with
the pin size equal to the interpin distance.  In this case the only possible
directions for orientational pinning are the [10] and [01], as we have seen
here.

It is worth noting that many experiments in JJA in the past have been done
in samples with a diagonal bias.  For example, van Wees {\it et
al.} \cite{expff}   have observed the existence of a transverse voltage in their
measurements, which was unexplained. 
From our finding that the [11]-direction is unstable against changes in
the angle of the bias, we conclude that any small deviation in the
direction of the flow of current, either due to tiny fabrication defects in the
busbars or to disorder in the critical currents of the junctions \cite{jlt},
may explain their observation.
Also Chen {\it et al.}\cite{chen} have reported a
transverse angle in measurements in JJA driven in the diagonal direction. In
their case the effect is antisymmetric against a
change in the direction of the magnetic field. 
Since transverse voltages due to the instability of the [11] direction
are even with the direction the magnetic field, 
their observation can not be explained from our results.
This means  that they have a
Hall effect, possibly due to quantum fluctuations. 
However, they report that their
transverse voltage had also a component which was even with the magnetic
field (which was discounted in their computation of the Hall angle). This
particular spurious contribution can also be attributed to a small
deviation in the direction of the bias or to disorder effects.  
From this we conclude that in
order to study the Hall effect in JJA the most convenient choice would be a
current bias in the [10] direction where the effect of transverse voltages
at small deviations in the bias or disorder is minimum.

When our work was upon completion, new studies of the effect of the
orientation of the bias in driven square JJA have appeared. Fisher, Stroud
and Janin \cite{stroud99} have studied some of the effects of the direction
of current in a fully frustrated JJA ($f=1/2$) at $T=0$. In their case a
transverse critical critical current and the dynamics as a function of $I_x$
and $I_y$ has been described. Their results are in part complementary to our
work with a single vortex ($f=1/L^2$).  Yoon, Choi and Kim \cite{choi} find
differences in the IV characteristics of JJA at $f=0$ when comparing the
parallel current bias with the diagonal current bias. Their results are in
agreement with our results.

It is worth to mention that very recently an experiment in another systems with  periodic pinning potentials,
in superconductors with  a square antidots arrays, evidence of guided vortex motion was observed \cite{colo}.
They showed that the pinning landscape provided by the square antidot lattice influences the vortex motion, 
given place to an anisotropic motion, temperature dependent, confirming our results
 of Sec.~1.3.5.  

In summary, in Sec.~1.3.5 of this review  we have considered  the dynamics of a single vortex in a
square JJA and diluted vortex lattices.  We were able to characterize in detail the orientational
pinning and breaking of rotational symmetry in this case. Furthermore, with
the results of the RSJ numerical calculation we were able to reproduce and
interpret most of  experimental measurements for a quasi-diagonal bias \cite{prbnos}. 
Regarding 
the moving vortex lattice different dynamical phases in a JJA,  as a
function of temperature and current, we  also show here that characteristics
 of the breaking of rotational invariance, orientational
pinning and transverse voltages depend on the dynamical phase under
consideration as well as on the disorder in the Josephson couplings \cite{jlt}.
Therefore we shown that orientational pinning is a useful tool to characterize
the non equilibrium vortex dynamics regimes in JJA, both numerically and experimentally.

\section{VORTEX DYNAMICS IN FULLY \\
FRUSTRATED JJA}

In this section we show the study of  the fully frustrated JJA  ($f=1/2$) for system
sizes of $L\times L$ junctions, with $L=8,16,24,32,48,64,128$. 
In the absence of external currents, we find an equilibrium
phase transition at $T_c=0.45$ which, within a resolution
of $\Delta T=0.005$,
corresponds to a simultaneous (or very close) 
breaking of the $U(1)$  and  the $Z_2$ symmetries.
Here we will analyze the possible occurrence of these transitions 
as a function of temperature when the JJA is driven by
currents well above the zero temperature critical current 
$I > I_{c0}=(\sqrt{2}-1)I_0\approx 0.414I_0$. 
\begin{figure}[t]
\begin{center}
\begin{minipage}[h]{70mm}
\centerline{\includegraphics[width=7.0cm]{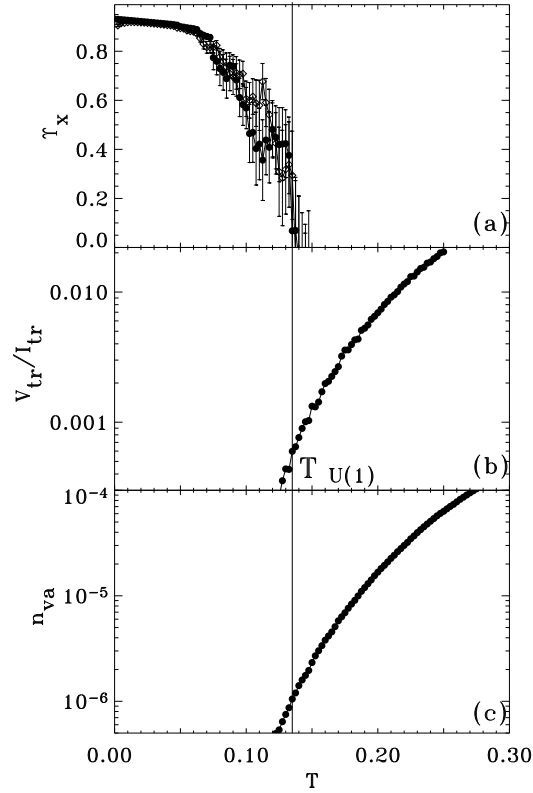}}
\end{minipage}
\caption{For large currents $I>I_c(0)$:
 breaking of the $U(1)$ symmetry.  Example showed for
$I=0.9$ and  system size $64\times64$. 
(a) Helicity Modulus $\Upsilon_x$ vs temperature $T$
($\bullet$ increasing $T$, $\Diamond$ decreasing $T$).
(b) Transverse voltage for a 
small transverse current, $I_{tr}=0.1$, vs $T$. 
(c) Vortex-antivortex pairs density, $n_{va}$ vs. $T$.}
\label{fig23}
\end{center}
\end{figure}

\subsection{$U(1)$ symmetry and transverse superconductivity}

In the driven JJA superconducting coherence can only be defined
in the direction transverse to the bias current \cite{prl1,prb2,kim93}.  
We calculate the transverse helicity modulus  as was shown in Sec.~1.2.3. 
We find that $\Upsilon_x$ is finite at low $T$  and vanishes
at a temperature $T_{U(1)}(I)$. In Fig.\ref{fig23}(a) we show the behavior
of $\Upsilon_x(T)$ for a current $I=0.9$ in a $64\times64$ JJA. 
This transition is reversible: we obtain the same behavior when
decreasing $T$ from a random configuration at $T=1$ and when
increasing $T$ from an ordered state at $T=0$, see Fig.\ref{fig23}(a).
Transverse superconductivity can be measured 
when a small current $I_{tr}$ is applied perpendicular to the driving current:
we find a  vanishingly small transverse voltage $V_{tr}$ below
$T_{U(1)}(I)$, as we found before in \cite{prl1,prb2} for $f=1/25$. 
We obtain the voltage  in the $\mu$-direction as the time average 
$V_\mu=\langle d\alpha_\mu(t)/dt\rangle$ (normalized
by $R_ {N}I_0$); longitudinal voltage is $V=V_y$ and transverse
voltage is $V_{tr}=V_x$. In Fig.\ref{fig23}(b) we see that the transverse
resistance $V_{tr}/I_{tr}$ is negligibly small for $T<T_{U(1)}$ and
starts to rise near the transition. The equilibrium 
$U(1)$  transition (at $f=0$, $I=0$, Kosterlitz-Thouless) 
is characterized by the unbinding
of vortex-antivortex pairs above $T_c$. We calculate the density
$n_{va}$ of vortex-antivortex excitations above checkerboard vortex
configuration as $2n_{va} = \langle|b({\bf \tilde n})|\rangle-f$,
where the  vorticity
at the plaquette ${\bf \tilde n}$ (associated to the site ${\bf n}$)
is $b({\bf \tilde n})=-\Delta_\mu\times{\rm nint}[\theta_\mu({\bf
n})/2\pi]$. We see in Fig.\ref{fig23}(c) that $n_{va}$ rises near $T_{U(1)}$.
Moreover, the transverse resistivity above $T_{U(1)}$ is
$V_{tr}/I_{tr}\propto n_{va}$.

\begin{figure}[t]
\begin{center}
\begin{minipage}[h]{60mm}
\centerline{\includegraphics[width=6.0cm]{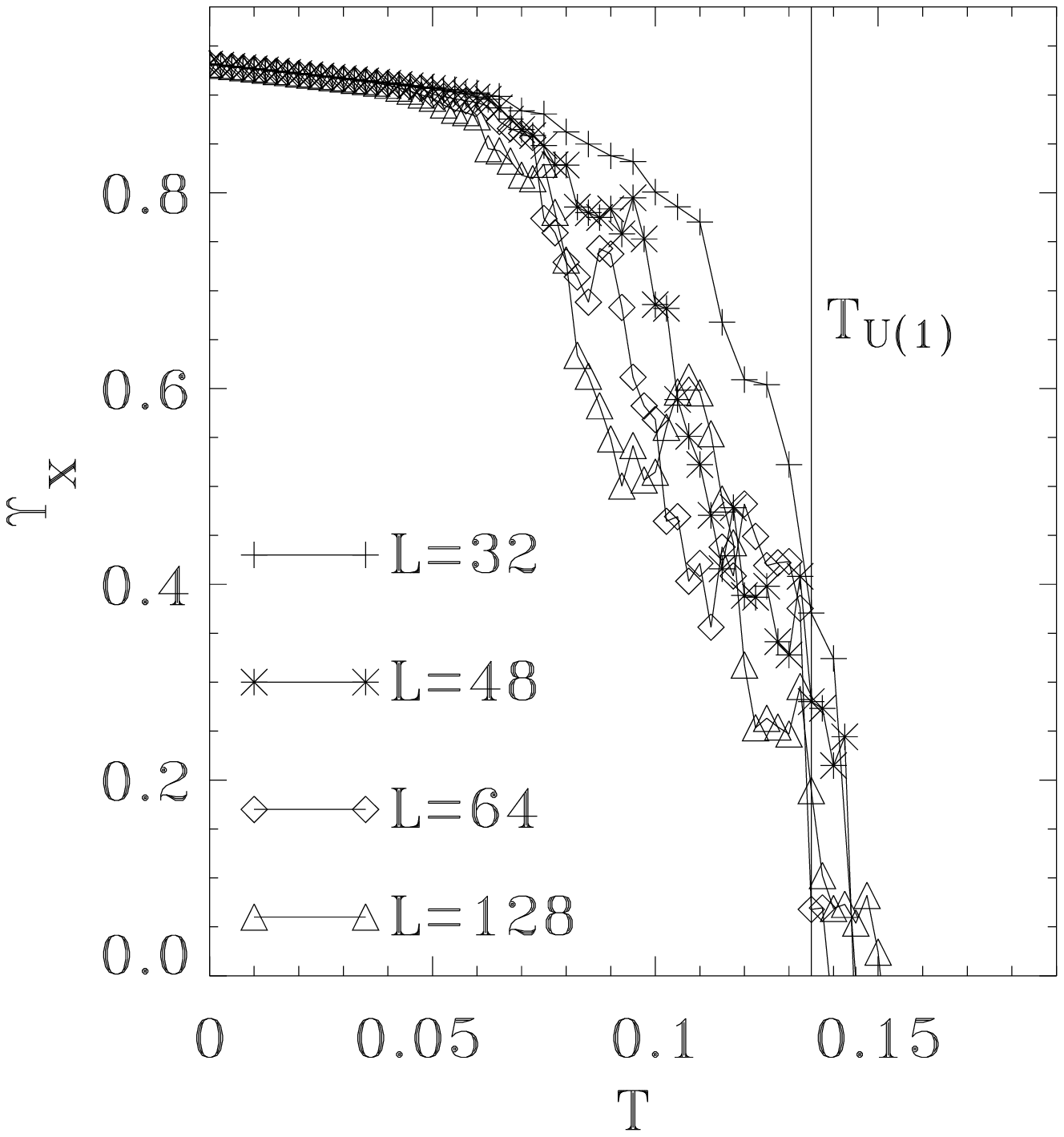}}
\end{minipage}
\caption{Helicity Modulus $\Upsilon_x$ vs temperature $T$
for $I=0.9$ and  system size $64\times64$, increasing $T$.
Size effect for $L=32,48,64,128$.}
\label{fig24}
\end{center}
\end{figure}

In Fig.\ref{fig24} shows $\Upsilon_x$ for sizes $L=32,48,64,128$,
we see that 
a transition temperature can be defined independently of lattice size.

\subsection{$Z_2$ symmetry}

\begin{figure}[t]
\begin{center}
\begin{minipage}[h]{65mm}
\centerline{\includegraphics[width=6.5cm]{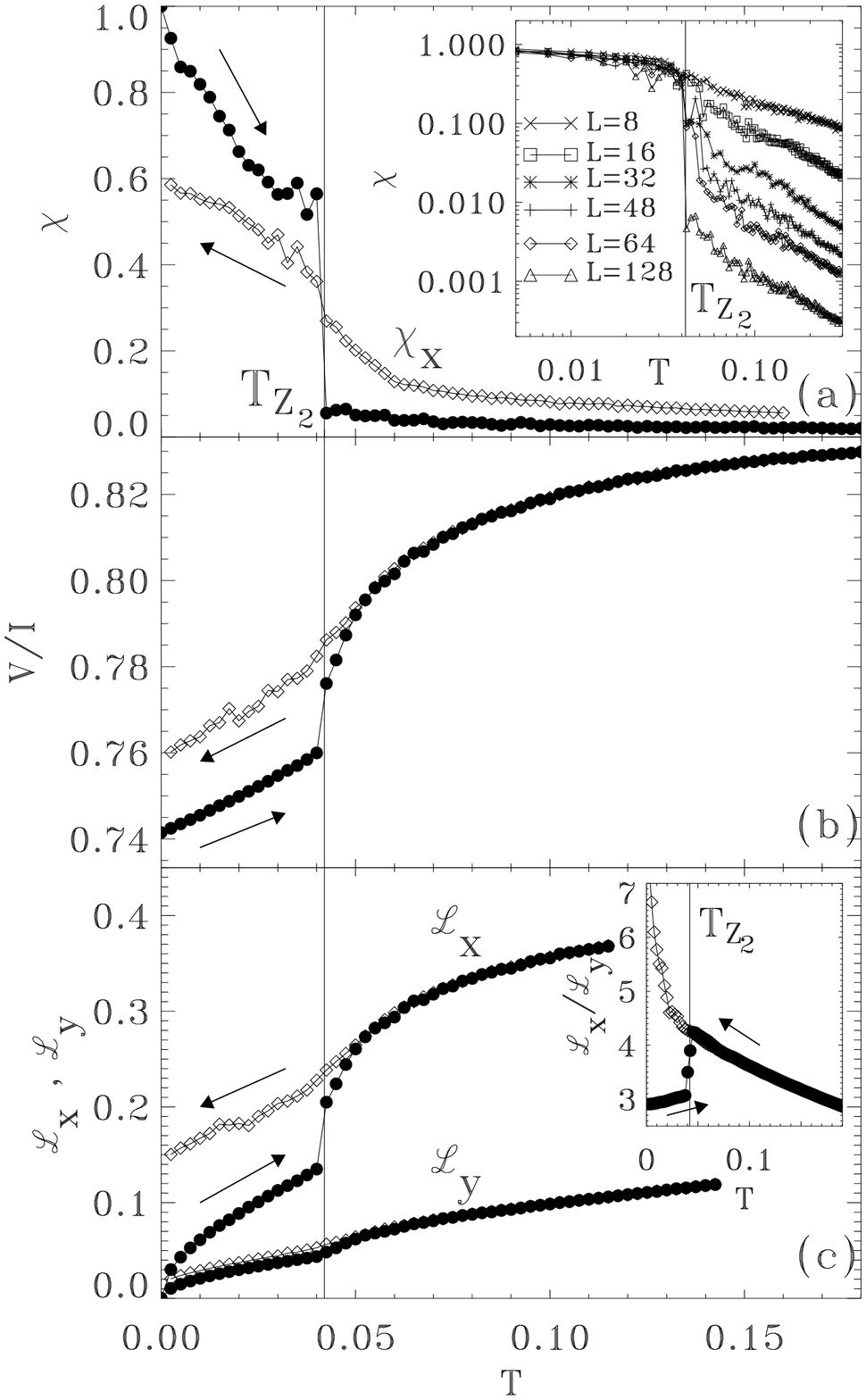}}
\end{minipage}
\caption{For a large currents, $I>I_c(0)$: 
breaking of the $Z_2$ symmetry.  Example shown for a dc current 
$I=0.9$ and system size $128\times128$.  Results
for  increasing $T$ ($\bullet$) and decresing $T$ ($\Diamond$).
(a) Chiral order parameter $\chi$ vs $T$ and one-dimensional
order parameter $\chi_x$ vs. $T$.
Inset: size effect for $\chi$ for $L=8,16,32,48,64,128$.
(b) Longitudinal voltage $V$ vs. $T$. 
(c) Domain wall lengths ${\cal L}_x$ and ${\cal L}_y$ vs. $T$.
Inset: domain anisotropy ${\cal L}_x/{\cal L}_y$ vs. $T$.}
\label{fig25}
\end{center}
\end{figure}
Since the ground state is a checkerboard pattern of vortices,
we define the ``staggered magnetization'' as
$m_s({\bf \tilde n},t)= (-1)^{n_x+n_y}[2b(n_x,n_y,t)-1]$ and
 $m_s(t)= (1/L^2)\sum_{\bf \tilde n}
m_s({\bf \tilde n},t)$. At $T=0$, $I=0$
there are two degenerate configurations with $m_s=\pm 1$. Above the
$T=0$ critical current  $I_{c0}$ the checkerboard state moves as
a rigid structure and $m_s(t)$ changes sign periodically with time. 
Therefore we define the
chiral order parameter as $\chi = \langle m_s^2(t)\rangle$.
We start the simulation at $T=0$ with an ordered
checkerboard state driven by a current $I>I_{c0}$ and then 
we increase slowly the temperature. We obtain that the chirality
parameter vanishes at a temperature $T_{Z_2}$, which is smaller than
$T_{U(1)}$, as can be seen in Fig.\ref{fig25}(a) for $I=0.9$. This transition is
confirmed by the size analysis shown in the inset of Fig.\ref{fig25}(a): 
for $T<T_{Z_2}$ the chirality $\chi$ is independent of size, while
for  $T> T_{Z_2}$ we see that  $\chi\sim 1/L^2$.
As it is shown in Fig.\ref{fig25}(b), the longitudinal voltage $V$ has a sharp
increase at $T_{Z_2}$, which could be easily detected
experimentally.
The excitations that characterize the $Z_2$ transition are domain
walls that separate domains with different signs of 
$m_s$. The length of domain walls in the direction
$\mu$ is given by 
${\cal L}_\mu = (2/L^2)\sum_{\bf \tilde n} \langle b({\bf \tilde n})
b({\bf \tilde n}+{\bf \nu})\rangle$, with $\nu \perp \mu$. 
We find that for $I>0$ and $T>0$ the domains are anisotropic, with 
the domain walls being longer in
the direction perpendicular to the current (${\cal L}_x > {\cal L}_y$) and
the domain anisotropy ${\cal L}_x/{\cal L}_y$ increasing with  $I$. 
In Fig.\ref{fig25}(c) we show the dependence of ${\cal L}_\mu$ with temperature for
$I=0.9$. At $T=0$ there are no domain walls, since the initial
condition is the checkerboard state, and the domain wall length
grows with $T$, showing a sharp increase at $T_{Z_2}$.
The domain anisotropy ${\cal L}_x/{\cal L}_y$ is shown in the inset of Fig.\ref{fig25}(c):
it has a clear jump at the transition in $T_{Z_2}$ while 
for $T\gg T_{Z_2}$ the domains tend to be less anisotropic.
When decreasing temperature from a random configuration at $T=1$, an
important number of domain walls along the $x$ direction remain frozen
below $T_{Z_2}$: ${\cal L}_x$ tends to a finite value when
$T\rightarrow0$  and 
the domain anisotropy tends to diverge  when cooling down. 
This leads to a strong hysteresis in the voltage $V$ at $T_{Z_2}$
(see Fig.\ref{fig25}(b))  since the extra domain walls increase dissipation
\cite{teitel89,simkin98}.
This low $T$ state with frozen-in domain walls is ordered along the
$x$-direction (i.e. perpendicular to $I$) but is disordered along the $y$
direction which gives $\chi\approx0$.
We define the  $Z_2$ order parameter in the $x$ direction as 
$\chi_x=\langle(1/L)\sum_{n_y}[(1/L)\sum_{n_x}m_s(n_x,n_y,t)]^2\rangle$
and $\chi_y$ is defined analogously.
We see in Fig.\ref{fig25}(a) that, when cooling down from high $T$, 
$\chi_x$ vanishes as $\chi_x \sim 1/L$ for $T>T_{Z_2}$ (it has stronger size
effects than $\chi$) 
and becomes finite for $T<T_{Z_2}$, while $\chi_y\approx 0$ for any $T$.  
Therefore, depending on the history, there are two kinds of high current
steady states with broken $Z_2$ symmetry at low $T$. One state has mostly the checkerboard
structure ($\chi\not=0$) with  few very anisotropic domains. It can
be obtained experimentally by cooling down at zero drive 
and then increasing $I$.
The other steady state is ordered in the direction perpendicular to $I$
($\chi_x\not=0$, $\chi_y=0$) with several domain walls 
along the $x$ direction. 
These domain walls move in the direction parallel
to $I$ (via the motion of vortices perpendicular to $I$)
giving an additional dissipation.
This state can be obtained experimentally by cooling down with a fixed $I$.
\begin{figure}[t]
\begin{center}
\begin{minipage}[h]{80mm}
\centerline{\includegraphics[width=8.0cm]{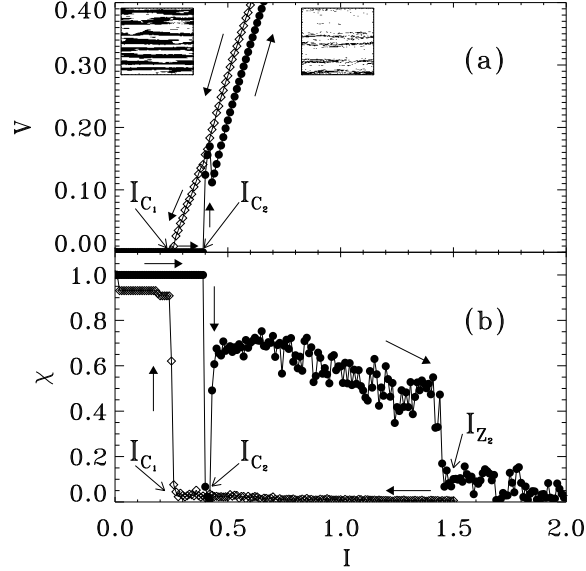}}
\end{minipage}
\caption{Current-voltage hysteresis for $T=0.02$ shown for 
(a) voltage vs. $I$ and 
(b) chiral order parameter vs.  $I$.
Increasing current from the checkerboard state ($\bullet$)
and decreasing current from a random state at large $I>I_{Z_2}$ 
($\Diamond$).
Insets: snapshots of the staggered magnetization $m_s({\bf n},t)$:
almost perfect ordered state obtained increasing current (right),
 and ordered state with ${\cal L}_x$-domain
walls obtained decreasing current (left).
}
\label{fig26}
\end{center}
\end{figure}

\begin{figure}[tbp]
\begin{center}
\begin{minipage}[h]{70mm}
\centerline{\includegraphics[width=7.0cm]{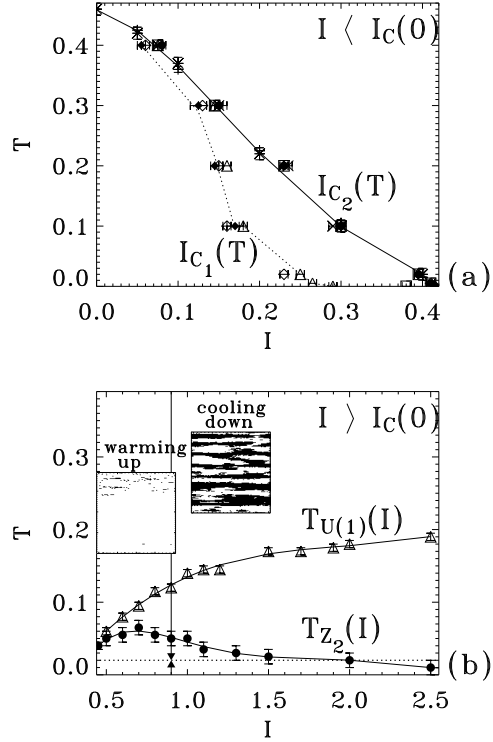}}
\end{minipage}
\caption{Current-temperature phase diagrams: (a) for low currents,
$I<I_c(0)$ and (b) for high currents, $I>I_c(0)$.
 $I_{c1}(T)$ ($\Diamond$) and $I_{c2}(T)$ ($\bullet$) lines
are obtained from hysteresis in IV curves as well as from hysteresis
in $\Upsilon_x(T)$ and $\chi(T)$ curves.
$T_{U(1)}(I)$  line obtained from $\Upsilon_x(T)$ and $V_{tr}(T)$
($\triangle$). $T_{Z_2}(I)$ line obtained from $\chi(T)$, $V(T)$ and
${\cal L}_x/{\cal L}_y(T)$ ($\bullet$).
The dashed line corresponds to the IV curve of previous Fig.
 ($T=0.02$). Insets: Two different steady states 
 (staggered magnetization snapshots) 
 found below $T_{Z_2}(I)$ line, at $T=0.02$ and $I=0.9$: (i) 
 warming up, almost a perfect checkerboard pattern,
 or (ii) cooling down, ordered state with ${\cal L}_x$-domain
walls.
}
\label{fig27}
\end{center}
\end{figure}

The two steady states have also different critical currents as can be
observed in the low $T$ current-voltage (IV) characteristics. In Fig.\ref{fig26}(a)
we show the IV curve for $T=0.02$ and in Fig~\ref{fig26}(b) the corresponding $\chi$
vs. $I$ curve. When increasing $I$ from the $I=0$ equilibrium state,
we find a critical current $I_{c2}(T)$, which in the limit of $T=0$
tends to $I_{c0}=\sqrt{2}-1$ as found analitically and in simulations
with PBC \cite{teitel83,rzch90,km}. 
Near $I_{c2}$ the order parameter $\chi$ has a minimum 
and rapidly increases with $I$. The driven state is an ordered state
similar to the one shown in Fig.\ref{fig25}(a) (see right inset).  At a higher current $I_{Z_2}$ there is 
a sharp drop of $\chi$ which corresponds to the crossing of the $T_{Z_2}(I)$
line (see Fig.\ref{fig27}) and the $Z_2$ order is lost.
If we now decrease the current either 
from the disordered state at $I>I_{Z_2}$
or from a random initial configuration  or from a configuration 
cooled down at a fixed $I > I_{c2}$,  we obtain the steady state with
domain walls along the $x$ direction and $\chi\approx0, \chi_x\not=0$ (see left inset).
This state has a higher voltage and pins at a lower critical current 
$I_{c1}(T)$, which has the $T=0$ limit $I_{c1}(T\rightarrow0)=0.35$.
It has been shown \cite{km} that open boundary conditions can nucleate domain
walls leading to the critical current $I_{c1}(0)=0.35$ usually found
in open boundary $T=0$ simulations \cite{teitel89,dyna0,falo,simkin98}. 
Also a moving state 
with parallel domain walls
(as in the insets of Fig.\ref{fig26}(a)) has been found by Gr\o nbech-Jensen {\it et
al.} \cite{niels} in $f=1/2$ JJA with open boundaries and 
Marino and Halsey \cite{marino} have 
shown that the high current states of frustrated JJA can have moving
domain walls. We have studied the effect of open boundaries 
in the direction of $I$, in the direction perpendicular to $I$ and in
both directions. They differ mainly in the $T=0$ critical current
and IV curve, for finite $T$ there are small differences 
in the detailed shape of the hysteresis in critical
current. In all the cases the two high current steady states are
observed at finite $T$ with the same history dependence.  
Also, we find that the density of frozen 
${\cal L}_x$ domain walls depends on cooling rate and decreases with
system size.

\subsection{Summary}

In summary, we have obtained the current-temperature 
phase diagram of the driven fully frustated XY model, which is shown in
Fig.\ref{fig27}(a) for low currents and in Fig.\ref{fig27}(b) for high currents.
 At high currents the breaking of the $U(1)$ and the $Z_2$
symmetries occurs at well separated temperatures, with
$T_{Z_2}<T_{U(1)}$. The low temperature regime $T<T_{Z_2}(I)$ has
bistability with two possible 
steady states (see insets in Fig.\ref{fig27}) and history dependent IV curves. 
The different transitions could  be observed
experimentally with measurements of the transverse and longitudinal
voltage.

\section{CONCLUSIONS}

 We presented a  detailed review  of our previous numerical studies on
non equilibrium  vortex dynamics in JJA
driven by a   dc current. Dynamical phase diagrams for different
magnetic field corresponding to  $f=1/25$ and $f=1/2$, current directions (orientational pinning)
 and  varying temperature were obtained and shown in Fig.\ref{fig23a} and Fig.\ref{fig27}.

In conclusion for diluted vortex lattices
we find that for low currents the depinning transition of the
VL and the later melting of the moving VL become separated with
$T_M(I)>T_p(I)$. For large currents we find that the moving VL has
a finite transverse critical current and therefore {\it transverse
superconducting coherence}. In this case it is possible to define
a transverse depinning transition at a temperature $T_{tr}$, and
a later melting transition of the moving VL at $T_M(I)$.
This transverse depinning transition could be easily studied
in controlled experiments in Josephson junction arrays, both with
transport measurements and with inductive coil measurements of the
transverse helicity modulus. This proposed experiment were later performed 
\cite{diego1} and  
the anisotropy of moving vortex phases was confirmed.

Regarding fully frustrated JJA  we obtain the current-temperature 
phase diagram. At high currents (Fig.\ref{fig27}(b)) we find
$T_{Z_2}<T_{U(1)}$, which is in contrast with the equilibrium result
of $T_{Z_2}\simeq T_{U(1)}$ [1,5].
Interestingly, the frustrated  XY model with modulated anisotropic couplings
also has $T_{Z_2}<T_{U(1)}$ \cite{berge86}, meaning that the anisotropy
induced by the current may provide a similar effect.
It is clear that the $f=1/2$ case has a strong pinning
effect (with $I_{c0}=0.414$) when compared
to the dilute case of $f=1/25$ (with $I_{c0}=0.11$) \cite{prl1,prb2}.
In fact, the transverse depinning temperature $T_{U(1)}$ is
one order of magnitude higher for $f=1/2$ with respects to
 $f=1/25$ \cite{prl1}.
The driving current weakens the effect of pinning
and thus $T_{U(1)}$ increases with $I$. A similar effect
gives a $T_{Z_2}$ growing with $I$ just above $I_{c0}$.
However for larger currents (near the Josephson current $I_0$) 
$T_{Z_2}(I)$ starts to decrease with $I$, 
with the limit $T_{Z_2}(I\rightarrow\infty)\rightarrow0$.
This is because a driving current increases
the density of domain walls 
(an effect already mentioned in \cite{teitel89}) destroying
the Ising order for $I\gg I_{c0}$. 
Moreover, we find that the ordered region in $T<T_{Z_2}(I)$ has
bistability with two possible 
steady states and history dependent IV curves. 
The different transitions here characterized could  be obtained
experimentally by measurements of  transverse and longitudinal
voltage. With our numerical results 
   on highly driven  fully frustrated JJA we expect to  
   motivate an experimental confirmation 
 of the clear separation of the transition temperatures, 
$T_{Z_2}<T_{U(1)}$.

\section*{Appendix A: Periodic boundary conditions}

\subsection*{Phases}

We want to obtain the periodic boundary condition (PBC) 
for superconducting phases $\theta(n_x,n_y)$.
In general we can write the PBC as:
\begin{eqnarray}
\theta({\bf n}+{\bf L_x})&=&\theta({\bf n})+u_x({\bf n})\nonumber\\
\theta({\bf n}+{\bf L_y})&=&\theta({\bf n})+u_y({\bf n})\;,
\end{eqnarray}
where ${\bf L_x}=(L_x,0)$ and ${\bf L_y}=(0,L_y)$. 
Taking into account that all variables should be independent of the
order of subsequent global translations and that the phases are defined
except for an addition of $2\pi l$, 
we are lead to the consistency condition
\begin{equation}
u_x({\bf n})+u_y({\bf n}+{\bf L_x})=
u_x({\bf n}+{\bf L_y})+u_y({\bf n})+2\pi l\;,
\label{cc}
\end{equation}
Therefore, in order to specify the 
periodic boundary conditions we have to give the functions $u_x, u_y$, 
which will depend on the gauge for the vector potential ${\bf A}$.

The periodic boundary condition for the phases can be deduced
by requiring that all
physical quantities are invariant after a translation in the lattice
size. The supercurrents $S_\mu({\bf n})=
\sin[\Delta_\mu\theta({\bf n})-A_{\mu}({\bf n})]$ should satisfy:
\begin{eqnarray}
S_\mu({\bf n}+{\bf L_x})&=&S_\mu({\bf n})\nonumber\\
S_\mu({\bf n}+{\bf L_y})&=&S_\mu({\bf n})\;.
\end{eqnarray}
 This implies that the gauge invariant phase difference 
$\theta_\mu({\bf n})=\Delta_\mu\theta({\bf n})-A_{\mu}({\bf n})$ 
should satisfy
\begin{eqnarray}
\theta_\mu({\bf n}+{\bf L_x})&=&\theta_\mu({\bf n})+2\pi l\nonumber\\
\theta_\mu({\bf n}+{\bf L_y})&=&\theta_\mu({\bf n})+2\pi l'\;,
\end{eqnarray}
with $l, l'$ any integer. This condition leads to
\begin{eqnarray}
\Delta_\mu u_x({\bf n})&=&2\pi l 
+A_\mu({\bf n}+{\bf L_x})-A_\mu({\bf n})\nonumber\\
\Delta_\mu u_y({\bf n})&=&2\pi l' 
+A_\mu({\bf n}+{\bf L_y})-A_\mu({\bf n})\;.
\end{eqnarray}
We can choose the solution with $l=l'=0$.  
In the Landau gauge,  $A_x$ is a linear function of $n_y$ 
and $A_y$ is a linear function of $n_x$. 
Taking the origin such that $A_\mu({\bf n}=0)=0$, 
we obtain for the Landau gauge:
\begin{eqnarray}
 u_x({\bf n})&=&u_x(0)+A_y({\bf L_x})n_y\nonumber\\
 u_y({\bf n})&=&u_y(0)+A_x({\bf L_y})n_x\;.
\end{eqnarray}
The consistency condition (\ref{cc}) requires
\begin{equation}
A_y({\bf L_x})L_y-A_x({\bf L_y})L_x=2\pi l\;.
\label{cc2}
\end{equation}
The term in the left side is equal to the total flux $2\pi f L_xL_y$,
therefore (\ref{cc2}) is equivalent to flux quantization, giving  
$f L_x L_y=N_v$, with $N_v=l$   the number of vortices.

If we take  $u_x(0)=u_y(0)=0$, we obtain for the PBC 
\begin{eqnarray}
\theta(n_x+L_x,n_y)&=&\theta(n_x,n_y)+A_y({\bf L_x})n_y  \nonumber \\
\theta(n_x,n_y+L_y)&=&\theta(n_x,n_y)+A_x({\bf L_y})n_x.
\end{eqnarray}
A particular choice can be the  gauge with 
$A_x({\bf n})=-2\pi f n_y$, $A_y({\bf n})=0$, which leads to Eq.~\ref{pbcp}.

\subsubsection{External currents and electric fields}

In the presence of external currents or voltages the periodic boundary
conditions have to be reconsidered. 
In this case it is possible to have $\oint {\bf E}\cdot{\bf dl}\not =0$
in a path that encloses all the sample either in the $x$ or the $y$
direction. 
Therefore, in a closed path we have to consider the Faraday's law
$$ \oint {\bf E}\cdot{\bf dl} = -\frac{1}{c}\frac{d\Phi}{dt}.$$
The two-dimensional sample with PBC can be thought as
the surface of a torus in three-dimensions. The closed paths we
are considering are the two pathes that encircle the torus. 
The electric field is not a gradient of a potential, 
it is now given by ${\bf E}=-{\bf \nabla}V-\frac{1}{c}\frac{\partial {\bf
A}}{\partial t}$.  One possible solution is to consider a
vector potential $${\bf A}({\bf r},t)={\bf A_0}({\bf r})-{\bf\vec
\alpha}(t),$$ for which 
\begin{eqnarray}
{\bf H}_{\rm ext}&=&{\bf \nabla}\times{\bf A_0}\nonumber\\
{\bf E}_{\rm total}&=&\frac{1}{c}\frac{d{\bf\vec\alpha}}{dt}\nonumber
\end {eqnarray}

In our case, we take the adimensional vector potential as:
\begin{equation}
A_\mu({\bf n},t) = A_\mu^0({\bf n}) -\alpha_\mu(t),
\end{equation}
with $A_\mu^0({\bf n})$ in the Landau gauge  
($A_x^0({\bf n})=-2\pi f n_y$, $A_y^0({\bf n})=0$). Therefore 
the gauge invariant phase is:
$$\theta_\mu({\bf n},t)=\Delta_\mu\theta({\bf n},t)-A^0_\mu({\bf
n})+\alpha_\mu(t).$$
Then $\alpha_\mu$ acts as a global time-dependent phase in
the $\mu$ direction.

In the normalized units used in this paper, 
the electric field in the link defined by the junction ${\bf n}, \mu$ is 
$$E_\mu({\bf n})=-\Delta_\mu V({\bf n})- \frac{d A_\mu({\bf n})}{dt},$$
where the electrostatic potential is $V({\bf n})=-\frac{d\theta({\bf
n})}{dt}$. Therefore we have
$$E_\mu({\bf n})=\Delta_\mu\frac{d\theta({\bf
n})}{dt}+\frac{d\alpha_\mu}{dt}.$$
The average electric field in the $\mu$ direction is:
$$E^{\rm av}_\mu=\frac{1}{L_xL_y}\sum_{\bf n}E_\mu({\bf
n})=\frac{d\alpha_\mu}{dt},$$
where we have used the fact that $\sum_{\bf n}\Delta_\mu\frac{d\theta({\bf
n})}{dt}=0$
(which is the discrete equivalent of $\oint{\bf \nabla}V\cdot{\bf
dl}=0$).
The current in the link ${\bf n},\mu$ is, in normalized units,:
$$I_\mu({\bf n})=E_\mu({\bf n})+\tilde{S}_\mu({\bf n}),$$
with $\tilde{S}_\mu({\bf n})=S_\mu({\bf n})+\eta_\mu({\bf n},t)$.
Therefore the average current in the $\mu$ direction is
\begin{eqnarray}
I^{\rm av}_\mu&=&\frac{1}{L_xL_y}\sum_{\bf n}I_\mu({\bf n})\nonumber\\
&=&E^{\rm
av}_\mu+S^{\rm av}_\mu+\eta^{\rm av}_\mu=
\frac{d\alpha_\mu}{dt}+\frac{1}{L_xL_y}\sum_{\bf n}\tilde{S}_\mu({\bf n}) \nonumber
\end{eqnarray}
There are two cases to consider: (i) {\it external current source}:
 the  external current is given
and the total voltage fluctuates and (ii) {\it external voltage source}:
the  external voltage is given and the total current fluctuates.

{\it(i) External current source:} the average current in
$\mu$ direction is fixed by the external current:
$I^{\rm av}_\mu=I_{\mu}^{\rm ext}$. 
The average electric field is a fluctuating quantity given by:
\begin{equation}
E^{\rm av}_\mu(t)=\frac{d\alpha_\mu}{dt}=I^{\rm ext}_\mu-\frac{1}{L_xL_y}
\sum_{\bf n}\tilde{S}_\mu({\bf n}).
\label{ecs}
\end{equation}
In this case, the $\alpha_\mu(t)$ is a dynamical variable, and its
time evolution is given by Eq.~\ref{ecs}.

{\it (ii) External voltage source:}  the average electric field
in the $\mu$ direction is fixed by the external electric field:
$E^{\rm av}_\mu=E_{\mu}^{\rm ext}$. 
Therefore, now the $\alpha_\mu$ is
given by the external source:
$$\alpha_\mu(t)= \int E_{\mu}^{\rm ext} dt,$$
and $\alpha_\mu(t)= E_{\mu}^{\rm ext} t$, 
if $E_{\mu}^{\rm ext}$ is time-independent. The average current is
now a fluctuating quantity given by:
$$I^{\rm av}_\mu=E_{\mu}^{\rm ext}+\frac{1}{L_xL_y}\sum_{\bf n}\tilde{S}_\mu
({\bf n}).$$

Let us see how the PBC are affected by a change
of gauge. The gauge transformations are the following:
\begin{eqnarray}
\theta({\bf n})\rightarrow \theta'({\bf n})&=&\theta({\bf
n})+\phi({\bf n})\nonumber\\
A_\mu({\bf n})\rightarrow A'_\mu({\bf n})&=&A_\mu({\bf
n})+\Delta_\mu\phi({\bf n})\nonumber\\
V({\bf n})\rightarrow V'({\bf n})&=&V({\bf
n})-\frac{d\phi({\bf n})}{dt}\nonumber
\end{eqnarray}
An interesting choice is:
$$\phi({\bf n})=\vec\alpha\cdot{\bf n}=\sum_\mu\alpha_\mu n_\mu$$
In this gauge we have $\theta_\mu({\bf n})=\Delta_\mu\theta({\bf
n})-A^0_\mu({\bf n})$ and the PBC for the phases is:
\begin{eqnarray}
\theta({\bf n}+{\bf L_x})&=&\theta({\bf n})+A_y({\bf
L_x})n_y+\alpha_xL_x  \nonumber \\
\theta({\bf n}+{\bf L_y})&=&\theta({\bf n})+A_x({\bf
L_y})n_x+\alpha_yL_y,
\label{pbcp2}
\end{eqnarray}
and for the voltages:
\begin{equation}
V({\bf n}+{\bf L_\mu})=V({\bf n})-\frac{d\alpha_\mu}{dt}L_\mu=
V({\bf n})-E^{\rm av}_\mu L_\mu
\label{pbcv}
\end{equation}
The equations of motion in this gauge are:
\begin{eqnarray}
\frac{d\theta({\bf n})}{dt}&=&\sum_\mu\frac{d\alpha_\mu}{dt}
n_\mu-\sum_{\bf n'}G_{{\bf n},{\bf
n'}}\Delta_\mu\cdot\tilde{S}_\mu({\bf n'})\nonumber\\
\frac{d\alpha_\mu}{dt}&=&I^{\rm ext}_\mu-\frac{1}{L_xL_y}\sum_{\bf
n}\tilde{S}_\mu({\bf n})
\end{eqnarray}

The periodic boundary conditions with a fixed external current
using Eq.~\ref{ecs} was used previously in Ref. \cite{minnh} for
$f=0$ (it was called a ``fluctuating twist boundary condition'') and
in Ref.~\cite{dd99} for $f\not=0$.  Also, the periodic
boundary conditions in the gauge of Eqs.~\ref{pbcp2} and \ref{pbcv} were used
in Ref.~\cite{vicente} for a time dependent $d$-wave
Ginzburg-Landau model.

\subsubsection{Helicity modulus}

The helicitity modulus $\Upsilon_\mu$ expresses the ``rigidity'' of the
system with respect to an applied ``twist'' in the periodic boundary
conditions. The twist $k_\mu$ is defined as a phase change of $L_\mu k_\mu$
between the two opposite boundaries which are connected through the PBC
in the $\mu$ direction 
$$\theta({\bf n}+{\bf L_\mu})=\theta({\bf
n})+L_\mu k_\mu.$$
The helicity modulus is obtained from the free energy $F(T,k_\mu)$   as:
\begin{equation}
\Upsilon_\mu=\left.\frac{1}{L^2}\frac{\partial^2F(T,k_\mu)}{\partial
k_\mu^2}\right|_{k_\mu=0}
\end{equation}
It is clear from Eq.~\ref{pbcp2} that $k_\mu=\alpha_\mu$. Then, in order
to evaluate the helicity modulus, $\alpha_\mu$ must be set to zero.
This means that the helicity modulus can not be calculated in
the direction in which there is an applied current, since
it gives a fluctuating twist $k_\mu(t)$.

\section*{Appendix B: Algorithm}

The core of the numerical calculation is to invert the Eq.~\ref{dyn0}. 
This means to solve a discrete Poisson equation of the form,
\begin{equation}
\Delta_\mu^2f({\bf n})=d({\bf n})
\label{poiss}
\end{equation}
with the periodic boundary conditions
\begin{eqnarray}
f({\bf n}+{\bf L_x})&=&f({\bf n})\nonumber\\
f({\bf n}+{\bf L_y})&=&f({\bf n})\;.
\end{eqnarray}
The linear system of $L_xL_y$ equations of (\ref{poiss}) is singular.
Physically, the reason is that in Eq.(\ref{dyn0}),  $f({\bf
n})=d\theta({\bf n})/dt$ 
corresponds to a voltage, which is defined 
except for a constant.  We choose the voltage reference such
that it has zero mean:
\begin{equation}
\sum_{\bf n} f({\bf n})=0\;,
\end{equation}
other choices are also possible (like, for example, fixing $f({\bf
n_0})=0$ at  a given site ${\bf n}_0$).

The method we use to invert the Eq.~(\ref{poiss})  is based on 
the Fourier Accelerated and Cyclic Reduction (FACR) 
algorithm \cite{recipes}.
In this case, we take first the discrete Fourier transform
in the $x$ direction:
\begin{equation}
\tilde{d}(k_x,n_y)=\sum_{n_x} d(n_x,n_y)e^{i\frac{2\pi k_x n_x}{L_x}}\;,
\end{equation}
which with a fast Fourier algorithm takes a computation time of 
order $L_x\log L_x$.
This leads to the following equation:
\begin{equation}
\epsilon_{k_x}\tilde{f}(k_x,n_y)-
\tilde{f}(k_x,n_y-1)-\tilde{f}(k_x,n_y+1)=\tilde{d}(k_x,n_y)\;,
\label{fft}
\end{equation}
with $\epsilon_{k_x}=4-2\cos\frac{2\pi k_x}{L_x}$
and with boundary condition $\tilde{f}(k_x,L_y+1)=\tilde{f}(k_x,1)$.
This is a cyclic tridiagonal equation which can be solved with
a simple LU decomposition algorithm in a computation time
of order $L_y$ \cite{recipes}. In this way the 
$\tilde{f}(k_x,n_y)$ is obtained from (\ref{fft}). Finally we take
the inverse Fourier transform to obtain
\begin{equation}
f(n_x,n_y)=\frac{1}{L_x}\sum_{k_x} \tilde{f}(k_x,n_y)e^{-i\frac{2\pi k_x
n_x}{L_x}}\;.
\end{equation}
This algorithm takes a computation time which is of order
$L_xL_y(\log L_x + A)$ with the constant $A \sim 1 $. This is faster
than the two dimensional Fourier transform method used
by Eikmans and van Himbergen \cite{eik}, which takes a computation
time of order $L_xL_y(\log L_x + \log L_y)$.

\begin{center}
{\bf Acknowledgments}

We acknowledge fruitful discussions with H. Pastoriza, D. Shal\'om,
J.V. Jos\'e, E. Granato and useful comments that helped to get
 this final version, from P. Martinoli.
This work was supported by  CONICET,
ANPCYT PICT99-03-06343, 
CNEA P5-PID-93-7,
Fundacion Antorchas P14116-147 and Swiss National Science Foundation.

\end{center}

\end{document}